\documentclass[preprint,12pt]{elsarticle}
\usepackage{amsmath, amssymb, amsfonts, mathtools}
\usepackage{array}
\usepackage[caption=false,font=normalsize,labelfont=sf,textfont=sf]{subfig}
\usepackage{textcomp}
\usepackage{stfloats}
\usepackage{url}
\usepackage{verbatim}
\usepackage{graphicx}
\usepackage{tikz}

\usepackage[table]{xcolor}
\usepackage{colortbl}
\usepackage{bm}
\allowdisplaybreaks
\usepackage{enumitem}
\definecolor{Tdarkblue}{RGB}{0,45,98}   
\definecolor{lightblue}{RGB}{220,230,242}
\usepackage{xcolor}   
\usepackage{caption} 
\definecolor{darkblue}{RGB}{0,0,139}

\usepackage[skip=8pt]{caption}
\usepackage{mathtools}
\usepackage{algorithm}
\usepackage{algpseudocode}
\algrenewcommand\algorithmiccomment[1]{\hfill// #1}
\usepackage{booktabs,tabularx}

\newcolumntype{B}[1]{>{\bfseries\boldmath\arraybackslash}m{#1}}
\newcolumntype{C}{>{\centering\arraybackslash\bfseries\boldmath}c}

\usepackage{pifont} 
\newcommand{\cmark}{\ding{51}}
\newcommand{\xmark}{\ding{55}}
\usepackage[unicode,psdextra]{hyperref}
\begin{document}

\begin{frontmatter}

\title{Blockchain-Driven AI-Enhanced Post-Quantum Multivariate Identity-based Signature and Privacy-Preserving Data Aggregation Scheme for Fog-enabled Flying Ad-Hoc Networks}

\author[1]{Sufian Al majmaie\corref{cor1}}
\ead{al-majmaie.2@wright.edu}

\author[1]{Ghazal Ghajari}
\ead{ghajari.2@wright.edu}

\author[2]{Niraj Prasad Bhatta}
\ead{bhattand@mail.uc.edu}

\author[1]{Fathi Amsaad}
\ead{fathi.amsaad@wright.edu}

\ead{}

\cortext[cor1]{Corresponding author}

\address[1]{Department of Computer Engineering, Wright State University, Dayton, OH 45435, USA}
\address[2]{Department of Computer Engineering, University Of Cincinnati, Cincinnati, Ohio 45219, USA}

\begin{abstract}
The integration of Fog Computing with Flying Ad-Hoc Networks (FANETs) offers promising capabilities for decentralized, low-latency intelligence in UAV-based applications. However, the distributed nature, mobility, and resource constraints of FANETs expose them to significant security and privacy challenges, particularly against quantum threats. To address these issues, this work introduces a blockchain-based, AI-enhanced key management framework designed for fog-enabled FANETs. The proposed scheme employs a Post-Quantum Multivariate Identity-Based Signature Scheme (PQ-MISS) and Zero-Knowledge Proofs (ZKPs) to achieve secure key establishment, privacy-preserving data aggregation, and integrity verification. A polynomial composition-based encryption mechanism and an aggregate signature model support secure and efficient multi-device communication across fog and UAV layers. Fog servers construct partial blockchain blocks from validated UAV data. These blocks are completed and mined by Cloud Servers (CSs). AI algorithms then analyze the verified data to generate accurate predictions and insights. NS-3 simulations validate the efficiency of PQ-MISS in reducing communication overhead while improving the speed and reliability of data aggregation and verification. Comparative analysis demonstrates the proposed scheme's advantages over existing methods in computational cost, post-quantum security, and scalability, making it a robust solution for secure, intelligent, and future-ready FANET systems.
\end{abstract}


\end{frontmatter}

\section{Introduction}
Integrating Fog Computing with Flying Ad-Hoc Networks (FANETs) has emerged as a transformative approach for enabling low-latency, decentralized, and intelligent aerial applications. FANETs, consisting of unmanned aerial vehicles (UAVs) that communicate wirelessly with each other and with ground control stations, are increasingly employed in time-critical operations such as disaster response, border surveillance, agriculture, and real-time traffic monitoring. However, traditional cloud-based models suffer from high latency and single points of failure, making them unsuitable for the dynamic and distributed nature of FANET environments. Fog computing addresses these limitations by pushing computational resources closer to the edge, enabling localized decision-making, context-aware data aggregation, and reduced transmission overhead \cite{mosenia2016comprehensive, boyes2018industrial}.

Despite these advantages, FANETs remain highly susceptible to various security and privacy challenges due to their open communication medium, mobility, and limited computational resources. Attacks such as Sybil, impersonation, message replay, GPS spoofing, and eavesdropping are particularly problematic in UAV swarms \cite{zhang2014sybil, bernstein2017post}. Furthermore, secure authentication and data aggregation in such networks are constrained by latency requirements, energy limitations, and the need for real-time integrity verification. Existing cryptographic protocols such as RSA and ECC, while effective against classical threats, are vulnerable to quantum attacks as demonstrated by algorithms such as Shor's and Grover's. This has motivated research into post-quantum cryptography (PQC), particularly multivariate polynomial-based schemes that are both efficient and resistant to quantum adversaries \cite{ding2005rainbow, sakumoto2011provable}.

Multivariate public key cryptosystems (MPKCs), such as Unbalanced Oil and Vinegar (UOV) and Rainbow, have shown promise for post-quantum digital signatures due to their small key sizes and fast signing operations \cite{chen2019identity}. However, many of these schemes suffer from high verification costs, unproven resistance to chosen-message attacks, or limited scalability. Identity-based signature (IBS) schemes offer certificate-free public key infrastructures that are well suited for mobile networks such as FANETs. Notable studies, such as \cite{shen2013ibuov, luyen2019improved}, have explored IBS for multivariate settings, but these schemes were either insecure under adaptive attacks or computationally inefficient for aggregate verification.

In parallel, blockchain technology has emerged as a robust platform for decentralized authentication, secure data logging, and integrity assurance in distributed systems. Its application to FANETs has been explored in contexts such as flight path validation, access control, and UAV swarm coordination \cite{sakumoto2011provable, dorri2017blockchain}. However, conventional blockchain frameworks face challenges such as high latency and energy consumption when deployed in resource-constrained environments. Lightweight blockchain adaptations, including partial block construction at fog nodes and optimized consensus algorithms such as PBFT, have been proposed to mitigate these issues \cite{bera2022private, yu2022slap}.

Artificial intelligence (AI) also plays an increasingly important role in the FANET security stack. The large volume of heterogeneous data collected by UAVs, covering telemetry, imagery, environmental readings, and system metrics, offers significant potential for predictive analytics and anomaly detection. Deep learning models, such as Long Short-Term Memory (LSTM) networks, can forecast UAV trajectories, detect intrusions, and optimize routing paths \cite{salem2024advancing, okdem2024artificial}. When integrated with blockchain, AI algorithms can securely access and learn from verified data, improving threat intelligence and operational reliability without compromising privacy \cite{ullah2024aicyber}. Nevertheless, most existing studies treat AI, blockchain, and cryptographic security as isolated layers, missing the opportunity for a holistic co-design.

\sloppy 
Despite these advancements, no current solution provides an integrated framework that combines post-quantum identity-based authentication, blockchain-backed secure data aggregation, and AI-enabled predictive analytics tailored for the constraints of fog-enabled FANETs. Existing blockchain-UAV systems lack post-quantum resistance. AI-based UAV optimization frameworks overlook secure key management. Many PQC schemes are not designed to support multi-signer aggregation, leading to inefficiencies in resource-constrained aerial swarms.

To address this gap, we propose PQ-MISS, a blockchain-driven, AI-enhanced post-quantum multivariate identity-based signature and privacy-preserving data aggregation mechanism for fog-enabled FANET environments. This research builds upon existing works in identity-based cryptographic methods, blockchain security, and AI-driven enhancements for UAV networks. IBS schemes have been studied in multivariate cryptographic settings for securing IoT and ad hoc networks \cite{ding2005rainbow, sakumoto2011provable}. The security of FANET communications has been addressed using various cryptographic primitives, including public key infrastructures and aggregate signature schemes \cite{shen2013ibuov, luyen2019improved}. Furthermore, blockchain integration with AI has been explored to enhance UAV operations, ensuring secure and efficient data aggregation and prediction modeling \cite{bera2022private, yu2022slap}. The main contributions of this work are as follows:

\begin{itemize}
    \item We propose an aggregate signature mechanism based on a Post-Quantum Multivariate Identity-Based Signature Scheme (PQ-MISS) tailored for fog-enabled FANET applications. In the proposed design, UAV drones and smart devices act as signers, while fog devices (FDs) serve as aggregators. Once the aggregated signatures are verified, the responsible cloud server (CS) within the peer-to-peer CS network generates and mines blocks using a voting-based consensus algorithm. These blocks are then incorporated into a public blockchain to ensure secure and tamper-evident data storage. Authenticated data stored in the blockchain is subsequently leveraged by machine learning (ML) algorithms to produce accurate predictive analytics.
    
    \item We implement and evaluate the proposed PQ-MISS using the NS-3 network simulator. The computational time required for generating and verifying both single and aggregate signatures is measured across FANET devices, FDs, and CSs.
    
    \item We perform a blockchain simulation to evaluate the computational time for block mining under different transaction volumes and block sizes. Results show that the proposed mechanism achieves improved efficiency and enhanced security compared to existing benchmark schemes.
\end{itemize}

\section{Related Work}
\label{sec:relatedwork}

\subsection{Post-Quantum Cryptography and Identity-Based Signature Schemes}
Quantum computing poses a serious threat to traditional cryptographic mechanisms, prompting the development of post-quantum cryptographic (PQC) solutions. Multivariate Public Key Cryptosystems (MPKCs) have been extensively studied for their quantum resistance and efficiency in resource-constrained environments \cite{ding2005rainbow}. Rainbow and Unbalanced Oil and Vinegar (UOV) schemes offer high-speed signature generation and verification, with recent research focusing on optimizing key sizes and improving resistance to chosen-message attacks \cite{sakumoto2011provable, hulsing2016mq}.

Identity-Based Signature (IBS) schemes mitigate key management overhead by deriving public keys from user identities, thereby eliminating the need for digital certificates. Several studies have combined IBS with MPKCs to create post-quantum secure identity frameworks \cite{paterson2006efficient, luyen2019improved}. Shen et al. \cite{shen2013ibuov} proposed an identity-based scheme using UOV, but subsequent work by Luyen et al. \cite{luyen2019improved} revealed vulnerabilities stemming from improper parameter selection. Chen et al. \cite{chen2019identity} improved the structure by introducing a multivariate IBS with better efficiency and enhanced security features.

Despite these advancements, most existing schemes do not provide signature aggregation, certificate elimination, and low computational cost simultaneously—requirements that are essential for UAV-based applications. Our proposed PQ-MISS enhances unforgeability, removes the need for certificates, and supports aggregate signatures, making it more suitable for FANET deployment.

\subsection{Blockchain for Secure FANET Communications}
Blockchain has become a key enabler for securing Internet of Drones (IoD) and FANET systems due to its distributed trust model, tamper resistance, and transparent verification. It has been used for identity management, access control, and secure communication in UAV systems \cite{bera2022private}.

Das et al. \cite{das2021igcacs} introduced iGCACS-IoD, a certificate-based authentication scheme using blockchain for secure access control. Similarly, Bera et al. \cite{bera2022private} proposed an authenticity scheme for UAVs in agricultural settings based on private blockchain technology. However, both approaches rely on classical cryptographic primitives, leaving them vulnerable to quantum attacks.

Yu et al. \cite{yu2022slap} presented SLAP-IoD, an authentication protocol integrating Physical Unclonable Functions (PUFs) to enhance hardware security through a lightweight blockchain framework. While effective in certain contexts, it does not provide post-quantum security or support aggregate verification.

In contrast, our solution uses blockchain not only for logging but also for secure multi-party aggregation and validated UAV transactions, combined with a PQC-ready identity framework.

\subsection{AI-Driven Optimization for FANETs}
Machine learning (ML) and artificial intelligence (AI) are increasingly vital for enhancing autonomy, resource management, and threat analysis in UAV networks. FANETs generate large volumes of telemetry, mission data, and environmental information, which can be exploited by ML for route optimization, anomaly detection, and behavior prediction.

Recent works employ deep learning models such as Long Short-Term Memory (LSTM) networks to optimize networks and predict trajectories in dynamic UAV topologies \cite{zhang2022recurrent}. AI models trained on blockchain-verified datasets improve resilience to pattern-based and spoofing attacks while enabling real-time prediction \cite{bera2022private}.

Most existing approaches, however, apply AI in isolation, without embedding it into cryptographic or blockchain security layers. Our proposed framework tightly integrates AI with secure, blockchain-stored data to provide trustworthy, intelligent decision-making in adversarial environments.

\subsection{Comparative Analysis of Existing Schemes}
To position the proposed PQ-MISS framework in the broader context, it is needed to survey state-of-the-art blockchain- and signature-based solutions for UAV, FANET, and IoT cases. The works exhibit significant progress in decentralized trust management, agile cryptography, and lightweight authentication, but they mostly suffer from limitations due to scalability constraints, vulnerability against quantum-capable adversaries, or restricted adaptability for highly dynamic FANET networks. For a systematic comparison, the comparison analysis in the remainder of this section identifies five key dimensions: post-quantum resistance (PQ), identity-based signature use (IBS), aggregate signature support (Agg.), blockchain support (BC), and the characteristic property describing each design. Table~\ref{tab:comparative-analysis} gives an overview of these properties among those of representative state-of-the-art solutions and the proposed PQ-MISS. As indicated by the table, the prior solutions normally meet only a few of these requirements, whereas PQ-MISS meets all of them by following an integrated construction optimized for scalable and secure FANET deployments.

\begin{table}[ht]
\scriptsize
\renewcommand{\arraystretch}{1.15}
\centering
\caption{Comparative analysis of PQ-MISS and existing post-quantum blockchain and signature schemes}
\label{tab:comparative-analysis}

\begin{tabular}{|m{2.5cm}|m{1cm}|m{1cm}|m{1cm}|m{1cm}|m{4.6cm}|}
\hline
\textbf{Scheme} & \textbf{PQ} & \textbf{IBS} & \textbf{Agg.} & \textbf{BC} & \textbf{Distinctive Feature} \\
\hline

\textbf{Proposed} & \cmark & \cmark & \cmark & \cmark & Fog-layer signature aggregation with AI-assisted post-quantum inference in FANETs \\
\hline
Holcomb et al.  \cite{holcomb2021pqfabric} & \cmark & \xmark & \xmark & \cmark & Crypto-agile hybrid signatures in Hyperledger Fabric; migration to PQ via OQS library \\
\hline
Kim et al.  \cite{kim2024post} & \cmark & \xmark & \xmark & \cmark & Falcon-based delegated consensus with verifiable randomness and low energy usage \\
\hline
Zhang et al. \cite{zhang2024post} & \cmark & \cmark & \xmark & \cmark & Lattice-based IBS with traceability, forward security, and layered IoT model integration \\
\hline
Prajapat et al. \cite{prajapat2024designing} & \cmark & \cmark & \xmark & \xmark & Quantum designated verifier signatures with OTP, source hiding, and non-repudiation \\
\hline
Dong et al. \cite{dong2023blockchain} & \xmark & \xmark & \xmark & \cmark & Blockchain-based UAV re-authentication via clustering, task migration, and redundancy \\
\hline
Shen et al. \cite{shen2013ibuov} & \cmark & \cmark & \xmark & \xmark & Unbalanced Oil and Vinegar IBS with provable security against existential forgery \\
\hline
Luyen et al. \cite{luyen2019improved} & \cmark & \cmark & \xmark & \xmark & Rainbow-based IBS with EUF-CMA security and compact key and signature sizes \\
\hline
Yu et al. \cite{yu2022slap} & \xmark & \xmark & \xmark & \xmark & Lightweight PUF-based authentication for secure UAV-device pairing \\
\hline
Das et al. \cite{das2021igcacs} & \xmark & \xmark & \xmark & \cmark & Blockchain-enabled certificate-based access control with group and role-aware policies \\
\hline
\end{tabular}
\end{table}

Shen et al. \cite{shen2013ibuov} proposed the Identity-Based Unbalanced Oil and Vinegar (IBUOV) scheme to eliminate the need for certificates in multivariate cryptography by deriving public keys directly from identity information. While provably secure with minimal key management, it inherits UOV's long public key drawback, limiting its practicality for UAV platforms. Luyen et al. \cite{luyen2019improved} advanced a Rainbow-based IBS achieving existential unforgeability against chosen-message attacks (EU-CMA) with shorter signatures than IBUOV, but still burdened by large public keys and computational cost.

\sloppy
Zhang et al. \cite{zhang2024post} developed a lattice-based IBS (PQ-IDS) leveraging bimodal Gaussian distributions and discrete Gaussian sampling to achieve post-quantum security, identity binding, and blockchain integration via a three-layer IoT architecture. Kim et al. \cite{kim2024post} introduced the Post-Quantum Delegated Proof of Luck (PQ-DPoL) consensus, embedding the NIST-approved Falcon signature into blockchain consensus for fairness and quantum resistance, albeit with trade-offs in signature size and verification time. Holcomb et al. \cite{holcomb2021pqfabric} extended Hyperledger Fabric with PQFabric, using Open Quantum Safe (OQS) algorithms for quantum-safe migration. While improving long-term security, performance is impacted by larger signatures and hashing overhead.

Prajapat et al. \cite{prajapat2024designing} presented a quantum designated verifier signature (QDVS) scheme relying on entangled quantum states, secure but impractical for UAVs without quantum infrastructure. Yu et al. \cite{yu2022slap} proposed the lightweight SLAP-IoD PUF-based authentication scheme, verified via AVISPA and Real-Or-Random (ROR) analysis, but lacking blockchain integration and multi-party signing.

Dong et al. \cite{dong2023blockchain} introduced a blockchain-based identity authentication framework tailored for multi-cluster UAV networking. Their design incorporates a UAV disconnection–reconnection mechanism and a task result backup strategy to enhance cluster robustness and preserve task integrity. While the approach effectively mitigates availability issues in UAV swarms, it relies on classical cryptographic primitives and omits support for post-quantum resilience and identity-based aggregation. Consequently, despite its strengths in reliability, the scheme remains unsuitable for long-term FANET security in the presence of quantum-capable adversaries.

Finally, Das et al. \cite{das2021igcacs} developed iGCACS-IoD, enhancing key management and preventing impersonation, yet relying on classical cryptography vulnerable to quantum adversaries.

This review underscores the progress in identity-based and post-quantum signature protocols, with multivariate and lattice-based techniques dominating. Nonetheless, lightweight designs with native blockchain integration remain essential for future-proof UAV and IoT security.
\subsection{Paper Organization}
The remainder of this paper is organized as follows. Section~\ref{sec:preliminaries} introduces notation and hardness assumptions, recalls the 5-pass identification protocol, formalizes the ISS model, and states the \textit{uf-cma} security game. Section~\ref{sec:networkmodel} presents the fog-enabled FANET architecture and trust assumptions. Section~\ref{sec:proposed} details the PQ-MISS design, including system setup and the KeyGen, LSSign, LSVer, LASign, and LAVer algorithms. Section~\ref{sec:application} shows the end-to-end application in IoT/FANETs, covering registration, data collection and fog-layer aggregation, blockchain block formation and PBFT-based commitment, AI-assisted prediction, and dynamic node onboarding. Section~\ref{sec:evaluation} reports NS-3.25 simulation results and blockchain-side computation costs, comparing PQ-MISS with MV-MSS \cite{srivastava2022blockchain} and LBAS \cite{bagchi2023public}. Finally, Section~\ref{sec:conclusion} concludes and outlines future directions.
\section{Preliminaries}
\label{sec:preliminaries}
Before describing the proposed PQ-MISS scheme, we outline the mathematical foundations and cryptographic tools on which our construction is built. This section introduces the underlying hardness assumptions, the multivariate identification protocol, identity-based signature framework, and the uf-cma security model used for our security analysis. Let $\hat{s}$ be the security factor, $x \in_R S$ where $x$ is selected randomly from set $S$, $\mathbb{F}_q$ denotes finite field containing $q$ elements, $\mathbb{F}_{q^\zeta}$ represent extension of $\mathbb{F}_q$ of degree $\zeta$, and $(\mathbb{F}_q)^\zeta = \{ y = (y_1, \ldots, y_\zeta) \mid y_i \in \mathbb{F}_q \}$, $i = 1, \ldots, \zeta$ represents vector space.

\subsection{Assumption related to Hardness}
The proposed MISS entrusts its security to the rigidity of the MQ problem which is represented as follows:
Given a system, 
$Q = (q_1^{(\rho_1, \ldots, \rho_n)}, \ldots, q_m^{(\rho_1, \ldots, \rho_n)})$ 
with $m$ quadratic equations involving $(\rho_1, \ldots, \rho_n)$ variables, it is essential to determine $n$ tuples $(\bar{\rho}_1, \ldots, \bar{\rho}_n)$ such that:
\[
\omega_1^{(\rho_1, \ldots, \rho_n)} = \cdots = \omega_m^{(\rho_1, \ldots, \rho_n)} = 0
\]

\subsection{Multivariate Identification Method}
\begin{figure}
    \centering
    \includegraphics[width=1\linewidth]{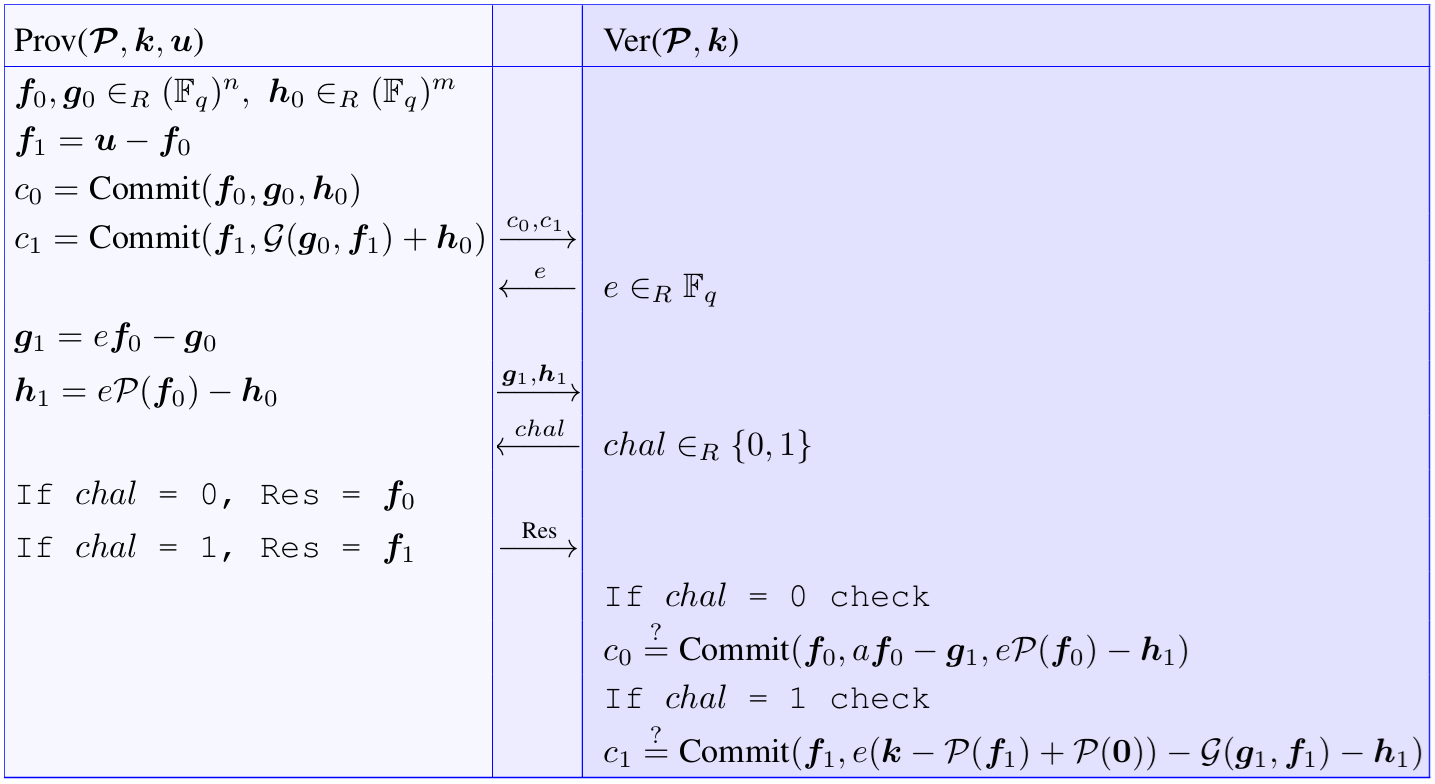}
    \caption{Sequence of steps in 5-Pass Identification Protocol}
    \label{fig1}
\end{figure}

Zero-knowledge (ZK) mechanisms allow clients to demonstrate that they possess specific information without revealing it to the verifier \cite{sakumoto2011provable}.For example, consider Alice who attempts to solve the Tower of Hanoi problem \cite{xu2001hm}. She wants to prove to Bob that she has the solution without disclosing it. She presents Bob with a specific instance of the problem and solves it using a modified approach. Once the solution is complete, she shows the solved puzzle to Bob. This helps Bob become convinced that she can solve the problem without revealing the actual solution method. The identification scheme can be built using the above idea and is shown in Figure~\ref{fig1}.

A MISS uses a randomly selected MQ system: $\mathbb{F}_q^n \rightarrow \mathbb{F}_q^m$. Security of a MISS depends on assumption that MQ problem seems to be hard. When Alice wants to show her ID to Bob, she is the prover, and Bob is the verifier who needs to be convinced. She desires to show Bob that she has understood the computation of $\tilde{s}$, a solution of $P(\tilde{x}) = \tilde{v}$ by not unveiling any information related to $\tilde{s}$. This mechanism may be built using ZKP. To build a ZKP with $\tilde{s}$, a function which is the polar form of $P$ given by
\[
G(V_1, V_2) = P(V_1 + V_2) - P(V_1) - P(V_2) + P(0)
\]
is defined. By using Commit, a computationally binding as well as statistically hiding commitment method, Sakumoto et al.~\cite{sakumoto2011provable} have built a 5-pass mechanism for identification using $\tilde{s}$. Assume that Soundness Error (SE), $SE = 1/2 + 1/2q$. To attain the anticipated security level and lessen the probability of impersonation, the protocol must be executed in several rounds. The number of rounds essential for reaching security level of $\zeta$ is
\[
r = \left\lceil \frac{-\zeta}{\log_2(KE)} \right\rceil
\]
\subsection{Identity-Based Signature Scheme (IBS)}
Identity-Based Signature (IBS) schemes allow users to generate digital signatures without relying on traditional certificate-based public key infrastructures. Instead, a trusted authority generates private keys using a user's unique identity (e.g., email, drone ID, etc.) as the public key input. This eliminates the need for complex key management systems, an advantage in resource-constrained environments such as FANETs.

We follow the IBS model introduced by Paterson and Schuldt \cite{paterson2006efficient}, which includes four main algorithms. Let $\hat{s}$ be the security parameter, $\text{ID} \in \{0,1\}^*$ a user identity, and $\text{Mg} \in \{0,1\}^*$ a message.

\begin{itemize}
    \item \textbf{Setup}$(1^{\hat{s}})$: Given the security parameter $\hat{s}$, the Secret Key Generation Center (KGC) generates the master public key $\text{Mpk}$ and master secret key $\text{Msk}$.
    \item \textbf{Extract}$(\text{Msk}, \text{ID})$: The KGC uses the master secret key $\text{Msk}$ and identity $\text{ID}$ to produce the user's signing key $U_{\text{ID}}$.
    \item \textbf{Signature Generation}$(U_{\text{ID}}, \text{Mg})$: Using $U_{\text{ID}}$ and message $\text{Mg}$, the user runs the signing algorithm to produce a signature $\varphi = \text{Sign}(\text{Mg})$.
    \item \textbf{Signature Verification}$(\text{Mpk}, \text{Mg}, \varphi, \text{ID})$: The verifier checks the validity of the signature $\varphi$ using the tuple $(\text{Mpk}, \text{Mg}, \varphi, \text{ID})$. If the signature is valid for the given identity, the algorithm returns $1$, otherwise it returns $0$.
\end{itemize}

\subsection{Unforgeability under Chosen-Message and Chosen-Identity Attack (uf-cma)}
The \textit{uf-cma} is a standard for ensuring security in an Identity-Based Signature Scheme, first introduced by \cite{goldwasser1988digital}. It is well-defined by an "experiment" or a "game" played between a Challenger ($\mathcal{C}$) and a Forger ($\mathcal{F}$). For an IBS with the same set of algorithms described earlier, the experiment $\text{Ex}_{\text{IBS}}(1^{\hat{s}})^{\text{uf-cma}}$ proceeds as follows: The Challenger ($\mathcal{C}$) executes $\text{Setup}(1^{\hat{s}})$ to generate the key pair $(\text{Mpk}, \text{Msk})$ and forwards $\text{Mpk}$ to the Forger ($\mathcal{F}$). The Forger sends queries dynamically to the Challenger.

\textbf{Extract-query:} For $\text{ID} \in \{0,1\}^*$, $\mathcal{F}$ asks $\mathcal{C}$ for the corresponding secret key. In response, $\mathcal{C}$ executes $\text{Extract}(\text{Msk}, \text{ID})$ to generate $U_{\text{ID}}$ and sends it to $\mathcal{F}$.

\textbf{Sign-query:} For $\text{ID} \in \{0,1\}^*$ and a message $\text{Mg}$, $\mathcal{F}$ requests a signature. In response, $\mathcal{C}$ first computes $U_{\text{ID}} = \text{Extract}(\text{Msk}, \text{ID})$, then generates $\varphi = \text{Sign}(\text{Mg})$, and sends it to $\mathcal{F}$.

Eventually, $\mathcal{F}$ outputs a tuple $(\text{ID}^*, \text{Mg}^*, \varphi^*)$. The Forger wins if:
\[
\text{Verify}(\text{Mpk}, \text{Mg}^*, \varphi^*, \text{ID}^*) = 1
\]
and $\mathcal{F}$ has not previously requested an \textit{Extract-query} on $\text{ID}^*$ or a \textit{Sign-query} on $(\text{ID}^*, \text{Mg}^*)$.

The success probability (or advantage) of the Forger is:
\[
\text{Adv}_{\mathcal{F}}^{\text{uf-cma}} = \Pr\left[\text{Ex}_{\text{IBS}}(1^{\hat{s}})^{\text{uf-cma}} = 1\right]
\]
An IBS is considered uf-cma secure if the probability above is negligible in $\hat{s}$.

\begin{figure*}[!t]
\centering
\includegraphics[width=1\linewidth]{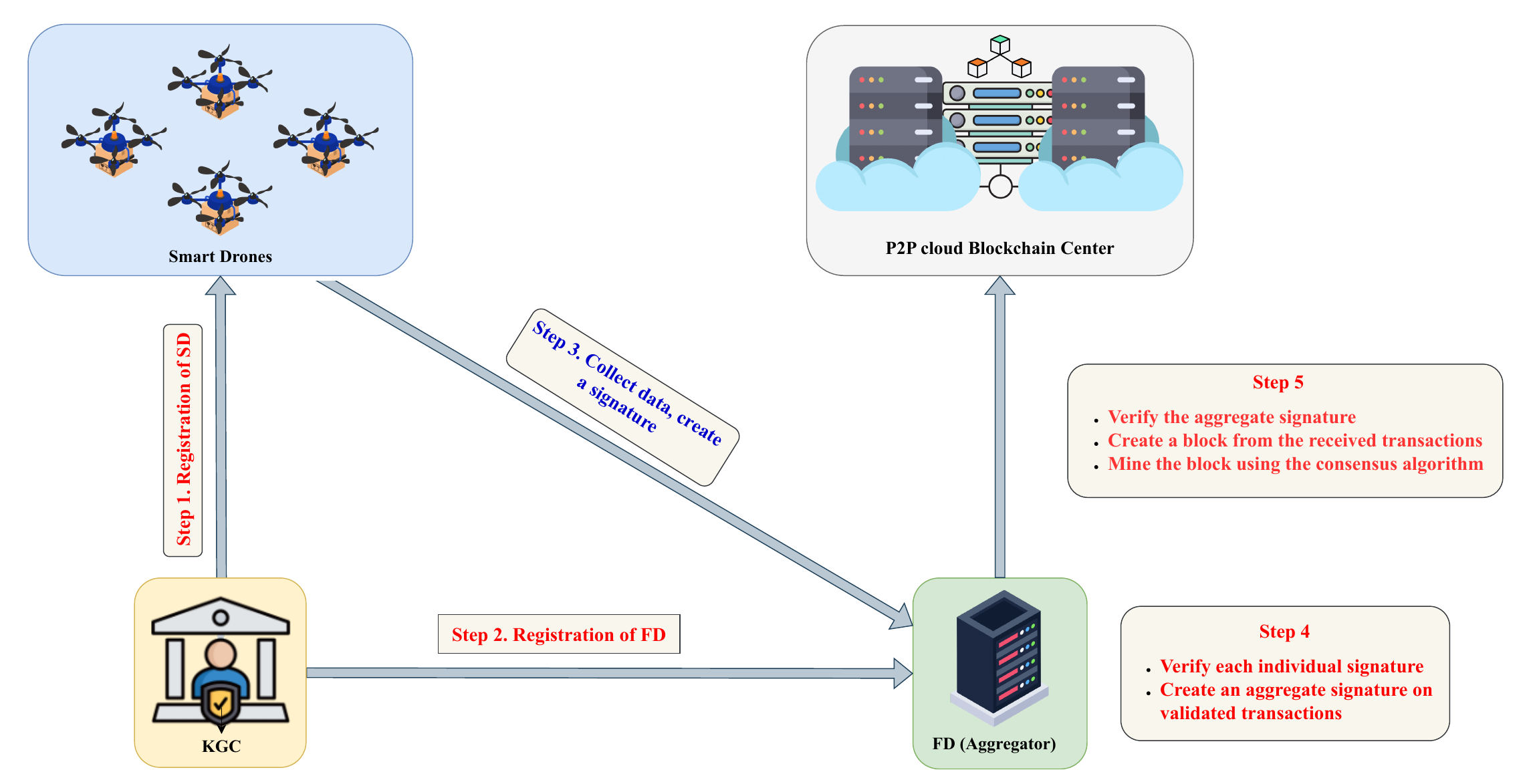}
\caption{Network model used in Fog-enabled Flying Ad-Hoc Networks}
\label{fig2}
\end{figure*}

\section{Network Model}
\label{sec:networkmodel}
In the proposed scheme, the network model includes drones, fog devices, and cloud servers as illustrated in Figure~\ref{fig2}. Our network model consists of the following components: drones, fog devices (FDs), a control room serving as the key generation center (KGC), and cloud servers. Drones are responsible for collecting data from designated application areas or flying zones and signing these messages. Each drone is equipped with IoT-enabled devices such as GPS, cameras, flight controllers, and transceivers.

The FD functions as an aggregator, gathering individually signed messages from its drones. It verifies each signature, creates an aggregate signature, and then sends this to a cloud server for storage in a blockchain network consisting of a Peer-to-Peer (P2P) cloud servers network. The network utilizes a consensus algorithm for block validation. Additionally, AI-based data analysis within the cloud servers utilizes genuine data stored in private blockchains to make accurate predictions.

The cloud server verifies the aggregate signature, creates blocks, and mines them to add to the public blockchain. The control room, acting as KGC, registers all drones and FDs by generating their key pairs using a lattice-based algorithm. The control room is fully trusted, while FDs are semi-trusted. Both are physically secured to prevent attacks. Cloud servers in the P2P network are also considered semi-trusted. After registration, drones and their corresponding FDs can be deployed operationally. The cloud servers, as part of a distributed system, facilitate the use of public blockchain storage.

\section{Proposed Scheme}
\label{sec:proposed}
\subsection{System Architecture Overview}
\label{subsec:architecture}
This work introduces a novel scheme named Post-Quantum Multivariate Identity-Based Signature Scheme (PQ-MISS), designed to resist quantum computing attacks in resource-constrained environments such as fog-enabled FANETs. The cryptographic foundations of PQ-MISS rely on the Multivariate Public Key Cryptosystem MISS proposed by Ding and Schmidt \cite{ding2005rainbow}, enhanced with the 5-pass identification protocol of Sakumoto et al. \cite{sakumoto2011provable} and transformed into a signature scheme using the method of H\"ulsing et al. \cite{hulsing2016mq}.

The PQ-MISS scheme consists of four algorithms:
\begin{itemize}
    \item \textbf{Setup:} The Key Generation Center (KGC) takes a security parameter $\hat{s}$ and generates a Master Public Key (Mpk) and a Master Secret Key (Msk). Mpk is made public while Msk remains private.
    \item \textbf{Extract:} KGC produces the private key of a user $(U_{\text{ID}})$ by implementing this algorithm for a user whose identity is $ID \in \{0,1\}^*$.
    \item \textbf{Signature Generation:} For message $Mg \in \{0,1\}^*$, a user with $ID \in \{0,1\}^*$ and $U_{\text{ID}}$ executes this procedure to generate signature $\varphi = \text{Sign}(Mg)$.
    \item \textbf{Signature Verification:} The correctness of the signature $(\varphi)$ is verified by the verifier by implementing this method on input that includes $Mpk$, $Mg$, $\varphi$, and $ID$.
\end{itemize}

The size of $Mpk$ is 
\[
\frac{m(n+2)(n+1)}{2}
\]
and the size of $Msk$ is 
\[
n^2 + m^2 + c
\]
field elements over $\mathbb{F}_q$, where $c$ represents the size of the central map, and $m$ and $n$ denote the number of public polynomials and variables of the PQ-MISS scheme, respectively. The parameters $m$ and $n$ in $\mathbb{F}_q$ elements also denote the sizes of user IDs and the private key, respectively. 

The size of the signature is 
\[
2\Psi \cdot |\text{Commitment}| + \Psi(m + 2n)
\]
$\mathbb{F}_q$ elements, where $\Psi$ represents the number of rounds required for the proposed PQ-MISS protocol. Luyen~\cite{luyen2019improved} stated that Shen et al.~\cite{shen2013ibuov} selected incorrect parameters for the security level and evaluated incorrect key sizes. In this work, parameters are chosen correctly. Furthermore, the ISS proposed by Shen et al.~\cite{shen2013ibuov} fails to provide \textit{uf-cma} security.

The IBS of Luyen~\cite{luyen2019improved} resembles a Public Key Infrastructure (PKI), where each user has a different public key (PK). The KGC links the PK of a user with their ID through a digital certificate, which is attached to a non-ID-based signature. In PQ-MISS, digital certificates are not required to associate a user's PK with their identity. 

In the IBS of Chen et al.~\cite{chen2019identity}, $\bar{F}$ and $Mpk$ are used as functions of $U_{ID} = (z_1, \ldots, z_d)$. If the signature generated by a user ($U_{ID}$) is to be verified, the expression of $\bar{F}$ for $U_{ID}$ must be evaluated, increasing the computation complexity for the verifier. The proposed scheme addresses this challenge since $Mpk$ is not a function of $U_{ID}$. 

The proposed scheme ensures \textit{uf-cma} security, similar to Luyen~\cite{luyen2019improved}. The IBS of Chen et al.~\cite{chen2019identity} does not ensure security. Our scheme also offers better performance than several prior works~\cite{mosenia2016comprehensive, luyen2019improved, chen2019identity} for the same sizes of $Mpk$, private key, and $Msk$, and involves a smaller signature size compared to some methods~\cite{shen2013ibuov, luyen2019improved}. The proposed scheme is highly suitable for a fog-based FANET network, where a lightweight IBS is essential for building a robust and secure system.

\subsection{Cryptographic Design of PQ-MISS}
\label{sec:design}
In the proposed PQ-MISS, MISS \cite{ding2005rainbow}, including 5-pass identification, is used as the fundamental block of the system. The scheme propounded by \cite{hulsing2016mq} is used in the proposed scheme. This method focuses on transforming the identification scheme into a signature scheme. PQ-MISS includes the above-mentioned algorithms. For input $\hat{s}$, KGC executes the \textit{Setup} procedure to generate $Mpk$ and $Msk$. During \textit{Extract}, KGC produces the private key of a user ($U_{ID}$) for a user with $ID \in \{0,1\}^*$. For $Mg \in \{0,1\}^*$, the \textit{Signature Generation} algorithm is executed by a user with $ID \in \{0,1\}^*$ and $U_{ID}$ to produce $\varphi = \text{Sign}(Mg)$. The verifier runs \textit{Signature Verification} on $(Mpk, Mg, \varphi, ID)$ to validate the signature correctness. \textit{Commit} is used in the proposed PQ-MISS.

\subsubsection{\texorpdfstring{Setup($1^{\hat{s}}$)}{Setup(1hs)}}

KGC executes \textit{KeyGen} on $1^{\hat{s}}$ for the MQ-based signature mechanism to produce 
\[
Mpk = P = S \circ F \circ T : \mathbb{F}_q^n \rightarrow \mathbb{F}_q^m, \quad Msk = \{S, F, T\}.
\]

\subsubsection{Extract(Msk, ID)}

For $Msk$ and $ID \in \{0,1\}^*$, KGC:
\begin{enumerate}
    \item Obtains $k_{ID} \in \mathbb{F}_q^m$ by determining $\text{Hash}(ID) = k_{ID}$, where $\text{Hash}()$ is a cryptographically secure and collision-resistant hash function.
    \item Determines $U_{ID} = P^{-1}(k_{ID}) \in \mathbb{F}_q^n$ using $Msk = \{S, F, T\}$.
    \item Forwards $U_{ID}$ to the user with identity $ID$.
\end{enumerate}

\subsubsection{\texorpdfstring{Signature Generation($U_{ID}$, $Mg$)}{Signature Generation(UID, Mg)}}

The user with $ID$ generates a signature for $Mg \in \{0,1\}^*$ with $U_{ID}$ as follows:

\begin{enumerate}
    \item Determines $a = \text{Hash}_1(P \| Mg)$ using $\text{Hash}_1$, a collision-resistant hash function that is cryptographically secure.
    \item Selects $f_0^1, \dots, f_0^{\Psi}, g_0^1, \dots, g_0^{\Psi} \in_R \mathbb{F}_q^n$ and $h_0^1, \dots, h_0^{\Psi} \in_R \mathbb{F}_q^m$, $j=1, \dots, \Psi$.
    \begin{enumerate}
        \item Computes $f_1^j = U_{ID} - f_0^j$.
        \item Assesses $\beta_0^j = \text{Commit}(f_0^j, g_0^j, h_0^j)$, $\beta_1^j = \text{Commit}(f_1^j, G(g_0^j, f_1^j) + h_0^j)$.
    \end{enumerate}
    \item Sets $\text{Comm} = (\beta_0^1, \beta_1^1, \dots, \beta_0^{\Psi}, \beta_1^{\Psi})$.
    \item Determines challenges $\text{Hash}_2(a \| \text{Comm}) = (\delta_1, \dots, \delta_{\Psi}) \in \mathbb{F}_q^{\Psi}$.
    \item Assesses $g_1^j = \delta_j f_0^j - g_0^j$ and $h_1^j = \delta_j P(f_0^j) - h_0^j$, $j=1, \dots, \Psi$.
    \item Writes $\text{Res}_1 = (g_1^1, h_1^1, \dots, g_1^{\Psi}, h_1^{\Psi})$.
    \item Determines challenge $\text{Hash}_3(a \| \text{Comm} \| \text{Res}_1) = (V_1, \dots, V_{\Psi}) \in \{0,1\}^{\Psi}$ using $\text{Hash}_3$.
    \item Sets $\text{Res}_2 = (f_{V_1}^1, \dots, f_{V_{\Psi}}^{\Psi})$.
    \item Returns signature as $\varphi = \text{Sign}(Mg) = (\text{Comm}, \text{Res}_1, \text{Res}_2)$.
\end{enumerate}

The length of the signature is $2\Psi \cdot |\text{Commitment}| + \Psi \cdot (m+2n)$ $\mathbb{F}_q$ elements.

\subsubsection{\texorpdfstring{Signature Verification($Mpk$, $Mg$, $\varphi = \text{Sign}(Mg)$, ID)}{Signature Verification(Mpk, Mg, varphi=Sign(Mg), ID)}}

The verifier performs the following steps to check the validity of $(Mg, \text{Sign}(Mg))$ for a user with $ID$:
\begin{enumerate}
    \item Determines $\text{Hash}(ID) = s_{ID}$.
    \item Evaluates $a = \text{Hash}_1(P \| Mg)$ and derives challenges $(\delta_1, \dots, \delta_{\Psi}) = \text{Hash}_2(a \| \text{Comm}) \in \mathbb{F}_q^{\Psi}$ and $(V_1, \dots, V_{\Psi}) = \text{Hash}_3(a \| \text{Comm} \| \text{Res}_1) \in \{0,1\}^{\Psi}$.
    \item Parses $\text{Comm}$ into $(\beta_0^1, \beta_1^1, \dots, \beta_0^{\Psi}, \beta_1^{\Psi})$, $\text{Res}_1$ into $(g_1^1, h_1^1, \dots, g_1^{\Psi}, h_1^{\Psi})$ and $\text{Res}_2$ into $(f_{V_1}^1, \dots, f_{V_{\Psi}}^{\Psi})$.
    \item For each $j = 1, \dots, \Psi$:
    \begin{itemize}
        \item If $V_j = 0$:
        \[
        \beta_0^j \stackrel{?}{=} \text{Commit}\left(f_{V_j}^j, \delta_j f_{V_j}^j - g_1^j, \delta_j P(f_{V_j}^j) - h_1^j \right)
        \]
        \item If $V_j = 1$:
        \[
        \beta_1^j \stackrel{?}{=} \text{Commit}\left(f_{V_j}^j, \delta_j (s_{ID} - P(f_{V_j}^j) + P(0)) - G(g_1^j, f_{V_j}^j) - h_1^j \right)
        \]
    \end{itemize}
\end{enumerate}
If all equalities hold, $(Mpk, Mg, \varphi, ID) = 1$; otherwise, $(Mpk, Mg, \varphi, ID) = 0$.

\textbf{Correctness:}
Correctness of MISS is proved by considering the equalities in Step~4.

For $V_j=0$,
\[
\beta_0^j=\text{Commit}\left(f_{V_j}^j,\;\delta_j f_{V_j}^j-g_1^j,\;\delta_j P(f_{V_j}^j)-h_1^j\right). \tag{1}
\]

For $V_j=1$,
\[
\beta_1^j=\text{Commit}\left(f_{V_j}^j,\;\delta_j(s_{\text{ID}}-P(f_{V_j}^j)+P(0))-G(g_1^j,f_{V_j}^j)-h_1^j\right). \tag{2}
\]

Where $j=1,\dots,\Psi$.

\noindent\textbf{Case 1:} If $V_j=0$, then $f_{V_j}^j = f_0^j$. 
$\delta_j f_{V_j}^j - g_1^j = \delta_j f_0^j - g_1^j = g_0^j$, and 
$\delta_j P(f_{V_j}^j) - h_1^j = \delta_j P(f_0^j) - h_1^j = h_0^j$. 
It is concluded that for $j=1, \dots, \Psi$, Eq.~(1) holds.

\medskip
\noindent\textbf{Case 2:} Let $V_j=1$. Then, $f_{V_j}^j = f_1^j$. Hence,
\begin{align*}
&\delta_j \big(s_{ID} - P(f_1^j) + P(0)\big) - G(g_1^j, f_1^j) - h_1^j \\
&= \delta_j \big(P(U_{ID}) - P(f_1^j) + P(0)\big) - G(g_1^j, f_1^j) - h_1^j \\
&= \delta_j \big(P(f_0^j + f_1^j) - P(f_1^j) + P(0)\big) - G(g_1^j, f_1^j) - h_1^j \quad \text{as } U_{ID} = f_0^j + f_1^j \\
&= \delta_j \big(P(f_0^j) + G(f_0^j, f_1^j)\big) - G(g_1^j, f_1^j) - h_1^j \\
&= G(\delta_j f_0^j - g_1^j, f_1^j) + \delta_j P(f_0^j) - h_1^j \\
&= G(g_0^j, f_1^j) + h_0^j
\end{align*}
Thus, for $j=1, \dots, \Psi$, Eq.~(2) holds.

\subsection{Post-Quantum Multivariate Digital Signature (PQ-MISS)}
\label{subsec:pqmiss}

This section introduces the Post-Quantum Verifiable Multivariate Identity-Based Signature Scheme (PQ-MISS), which enhances the standard MISS architecture to support aggregate signing and verification. The scheme is designed for use in distributed and constrained environments where multiple signers must produce verifiable multivariate signatures that can be compactly aggregated by a trusted aggregator (Agg).

The PQ-MISS scheme consists of five main algorithms: \textbf{KeyGen} for key generation of all users and the aggregator, \textbf{LSSign} for local signing by individual signers, \textbf{LSVer} for verification of each individual signature by the aggregator, \textbf{LASign} for aggregation of valid signatures into a compact group signature, and \textbf{LAVer} for final verification of the aggregate signature by an external verifier.

The system involves a Key Generation Center (KGC), $n$ individual signers, and a central aggregator (Agg). The KGC is a trusted authority that holds the master secret key (Msk) and master public key (Mpk). It is responsible for generating key pairs for all parties in the system and securely distributing them.

Each signer and the aggregator are assigned unique identities from the domain $\{0,1\}^*$, and their corresponding secret and public keys are derived using the multivariate structure
\[
P = S \circ F \circ T : \mathbb{F}_q^n \rightarrow \mathbb{F}_q^m,
\]
where $S$, $F$, and $T$ are affine and central transformations over $\mathbb{F}_q$. The KGC computes each user's secret key $U_{ID}$ as
\[
U_{ID} = P^{-1}(s_{ID}),
\]
where $s_{ID}$ is the hashed identity output using a one-way collision-resistant function.

In the following subsections, we describe the specific steps involved in each component of PQ-MISS. These include detailed procedures for key assignment, signer operations, multi-signature verification, and secure aggregate signature generation and validation. Notations with their descriptions are listed in Table~\ref{tab:symbols}.

\begin{table}[!t]
\caption{Symbol descriptions for the proposed PQ-MISS scheme}
\label{tab:symbols}
\centering
\renewcommand{\arraystretch}{1.3}
\setlength{\tabcolsep}{6pt}
\small
\begin{tabular}{|c|p{9cm}|}
\hline
\textbf{Symbol} & \textbf{Description} \\
\hline
$FD_j$ & $j^{\text{th}}$ Fog device \\
\hline
$SDp_j$ & $j^{\text{th}}$ FANET device group \\
\hline
$Agg$ & Aggregator \\
\hline
$DR_i$ & $i^{\text{th}}$ FANET device \\
\hline
$CS_k$ & $k^{\text{th}}$ semi-trusted cloud server \\
\hline
P2P CS & Peer-to-peer cloud server network \\
\hline
KGC & Key Generation Center (trusted control room) \\
\hline
$Msk$, $Mpk$ & Master secret/public key of KGC \\
\hline
$\mathbb{F}_q$ & Finite field with $q$ elements \\
\hline
$i$ & Integer in $\{1,2,\ldots,n\} \cup \{\text{ID}_{Agg}\}$ \\
\hline
$sk_i$, $pk_i$ & Secret/public key of $DR_i$ \\
\hline
$m_i$ & Message signed by $DR_i$ \\
\hline
$\sigma$ & Signature of $DR_i$ \\
\hline
$pk_{Agg_j}$ & Public key of aggregator \\
\hline
$sk_{Agg_j}$ & Secret key of aggregator \\
\hline
$m'$ & Message signed by aggregator \\
\hline
$ASig_{Agg_j}$ & Aggregate signature \\
\hline
$H(\cdot)$ & Collision-resistant hash function from $R_p$ to $R_p^1$ \\
\hline
$CTS_X$ & Current timestamp from $X$ \\
\hline
$\Delta T$ & Maximum permissible transmission delay \\
\hline
$\|$ & Data concatenation \\
\hline
$\hat{s}$ & Security parameter \\
\hline
$s_{ID}$ & Identity-derived value of signer \\
\hline
$s_{Agg}$ & Identity-derived value of aggregator \\
\hline
\end{tabular}
\normalsize
\end{table}

\subsubsection{KeyGen}

The Key Generation Center (KGC) performs the following steps to initialize the PQ-MISS system:

\begin{itemize}
  \item Selects a secret key $sk = U_{ID}$ and defines the corresponding public key as $pk = S \circ F \circ T : \mathbb{F}_q^n \rightarrow \mathbb{F}_q^m$.
  
  \item Chooses $(n+1)$ values from $\mathbb{F}_q^n$, denoted as $(s_1, s_2, \ldots, s_n)$ and $s_{ID}^{\text{Agg}}$, corresponding to the identities of each signer $i = 1, 2, \ldots, n$ and the aggregator $ID_{\text{Agg}}$.

  \item For each $i^{\text{th}}$ signer and the aggregator, where $i \in \{1, 2, \ldots, n\} \cup \{ID_{\text{Agg}}\}$, applies a one-way collision-resistant hash function $\bar{H} : \{0,1\}^* \rightarrow \mathbb{F}_q^m$ to compute $s_{ID} = \bar{H}(\text{Msk}, ID)$.

  \item Computes the corresponding private key for each participant as $U_{ID} = P^{-1}(s_{ID}) \in \mathbb{F}_q^n$, while the public key for all entities is $pk = S \circ F \circ T$.
\end{itemize}

As a result, each signer and the aggregator receives their own key pair $(sk_i, pk_i)$, derived from their identity and the system-wide parameters.

\subsubsection{LSSign}

Each signer $i$ holds a private–public key pair $(sk_i, pk_i)$. To sign a message $m_i \in \{0,1\}^*$, the signer executes the LSSign algorithm as follows.

First, the signer computes:
\[
c_i = \text{Hash}_1(pk_i \| m_i)
\]
using a cryptographically secure hash function.

For each $j = 1, \ldots, \Psi$, the signer randomly selects:
\[
f_0^j, g_0^j \in_R \mathbb{F}_q^n, \quad h_0^j \in_R \mathbb{F}_q^m.
\]

It then computes:
\[
f_1^j = U_{ID} - f_0^j
\]
and generates two commitments for each round:
\[
\beta_0^j = \text{Commit}(f_0^j, g_0^j, h_0^j), \quad
\beta_1^j = \text{Commit}(f_1^j, G(g_0^j, f_1^j) + h_0^j).
\]

The commitments are arranged as:
\[
\text{Comm} = (\beta_0^1, \beta_1^1, \ldots, \beta_0^{\Psi}, \beta_1^{\Psi}).
\]

A challenge vector is then computed:
\[
(\delta_1, \ldots, \delta_\Psi) = \text{Hash}_2(c_i \| \text{Comm}) \in \mathbb{F}_q^\Psi.
\]

For each $j = 1, \ldots, \Psi$, the responses are calculated:
\[
g_1^j = \delta_j f_0^j - g_0^j, \quad
h_1^j = \delta_j P(f_0^j) - h_0^j.
\]

These are collected into the first response sequence:
\[
\text{Res}_1 = (g_1^1, h_1^1, \ldots, g_1^{\Psi}, h_1^{\Psi}).
\]

A second challenge vector is generated as:
\[
(V_1, \ldots, V_\Psi) = \text{Hash}_3(c_i \| \text{Comm} \| \text{Res}_1) \in \{0,1\}^\Psi.
\]

The second response sequence is:
\[
\text{Res}_2 = (f_{V_1}^1, \ldots, f_{V_\Psi}^\Psi).
\]

Finally, the signer outputs the signature on $m_i$ as:
\[
\text{Sign}(m_i) = (\text{Comm}, \text{Res}_1, \text{Res}_2).
\]

\subsubsection{LSVer}

To verify a signature $\varphi = \text{Sign}(m_i)$ for a message $m_i$ from signer $i$, the aggregator (Agg) proceeds as follows.

It computes the identity hash $s_{ID} = \text{Hash}(ID)$ and the challenge seed $c_i = \text{Hash}_1(pk_i \| m_i)$. Based on these values, it derives the challenge vectors:
\begin{align}
(\delta_1, \ldots, \delta_\Psi) &= \text{Hash}_2(c_i \| \text{Comm}) \in \mathbb{F}_q^\Psi \\
(V_1, \ldots, V_\Psi) &= \text{Hash}_3(c_i \| \text{Comm} \| \text{Res}_1) \in \{0,1\}^\Psi
\end{align}

The signature $\varphi$ is parsed into:
\begin{itemize}
  \item $\text{Comm} = (\beta_0^1, \beta_1^1, \ldots, \beta_0^\Psi, \beta_1^\Psi)$
  \item $\text{Res}_1 = (g_1^1, h_1^1, \ldots, g_1^\Psi, h_1^\Psi)$
  \item $\text{Res}_2 = (f_{V_1}^1, \ldots, f_{V_\Psi}^\Psi)$
\end{itemize}

For each round $j = 1, \ldots, \Psi$, the aggregator checks the following verification condition:

If $V_j = 0$:
\[
\beta_0^j \stackrel{?}{=} \text{Commit}\left(
f_{V_j}^j,\,
\delta_j f_{V_j}^j - g_1^j,\,
\delta_j P(f_{V_j}^j) - h_1^j
\right)
\]

If $V_j = 1$:
{\small
\begin{align}
\beta_1^j \stackrel{?}{=} \text{Commit}\Big(& f_{V_j}^j,\; \delta_j (s_{ID} - P(f_{V_j}^j) + P(0)) \nonumber \\
& {} - G(g_1^j, f_{V_j}^j) - h_1^j \Big)
\end{align}
}

The signature is accepted if all such conditions hold for every round $j = 1, \ldots, \Psi$, i.e.,
\[
\text{Signature Verification}(Mpk, m_i, \varphi, ID) = 1
\]
Otherwise, the signature is rejected:
\[
\text{Signature Verification}(Mpk, m_i, \varphi, ID) = 0
\]

\subsubsection{LASign}

After verifying individual signatures $(\text{Sign}(m_i), ID_i)$ for $i = 1, 2, \ldots, n$, the aggregator (Agg) proceeds to generate a compact aggregate signature for the entire message set $(m_1, m_2, \ldots, m_n)$.

First, the aggregator computes $\hat{s}_{ID_i} = \text{Hash}(ID_i)$ for each signer. It then randomly selects values $f_0^{j,\text{Agg}}, g_0^{j,\text{Agg}} \in_R \mathbb{F}_q^n$, and $h_0^{j,\text{Agg}} \in_R \mathbb{F}_q^m$ for $j = 1, \ldots, \Psi$.

The commitments are generated as:
\[
\beta_0^{j,\text{Agg}} = \text{Commit}(f_0^{j,\text{Agg}}, g_0^{j,\text{Agg}}, h_0^{j,\text{Agg}})
\]
\[
\beta_1^{j,\text{Agg}} = \text{Commit}(f_1^{j,\text{Agg}}, G(g_0^{j,\text{Agg}}, f_1^{j,\text{Agg}}) + h_0^{j,\text{Agg}})
\]
where $f_1^{j,\text{Agg}} = U_{ID}^{\text{Agg}} - f_0^{j,\text{Agg}}$.

These are collected into:
\[
\text{Comm}^{\text{Agg}} = (\beta_0^{1,\text{Agg}}, \beta_1^{1,\text{Agg}}, \ldots, \beta_0^{\Psi,\text{Agg}}, \beta_1^{\Psi,\text{Agg}})
\]

Next, the challenge vector is derived as:
\[
(\delta_1^{\text{Agg}}, \ldots, \delta_\Psi^{\text{Agg}}) = \text{Hash}_2(c_i \| \text{Comm}^{\text{Agg}})
\]

For each round $j$, the aggregator computes:
\[
g_1^{j,\text{Agg}} = \delta_j^{\text{Agg}} f_0^{j,\text{Agg}} - g_0^{j,\text{Agg}}, \quad h_1^{j,\text{Agg}} = \delta_j^{\text{Agg}} P(f_0^{j,\text{Agg}}) - h_0^{j,\text{Agg}}
\]

The response vector is:
\[
\text{Res}_1^{\text{Agg}} = (g_1^{1,\text{Agg}}, h_1^{1,\text{Agg}}, \ldots, g_1^{\Psi,\text{Agg}}, h_1^{\Psi,\text{Agg}})
\]

A second challenge is computed:
\[
(V_1^{\text{Agg}}, \ldots, V_\Psi^{\text{Agg}}) = \text{Hash}_3(c_i \| \text{Comm}^{\text{Agg}} \| \text{Res}_1^{\text{Agg}})
\]

Then, the revealed responses are set as:
\[
\text{Res}_2^{\text{Agg}} = (f_{V_1}^{1,\text{Agg}}, \ldots, f_{V_\Psi}^{\Psi,\text{Agg}})
\]

The aggregator constructs the aggregate signature as $S_{\text{Agg}} = (\text{Sign}_i(m_i), ID_{\text{Agg}})$. If $S_{\text{Agg}} \notin P$, the process is restarted. Otherwise, the final aggregate signature is $(S_{\text{Agg}}, c_{\text{Agg}})$, which is transmitted to the verifier along with the individual components $\{(s_i, c_i, m_i)\}_{i=1}^n$.

\subsubsection{LAVer}

Upon receiving the aggregate signature $(S_{\text{Agg}}, c_{\text{Agg}})$ along with the supporting set $\{(s_i, c_i, m_i)\}_{i=1}^n$, the verifier executes the LAVer procedure to validate the authenticity of the aggregated result.

The verifier begins by parsing the aggregate signature components as follows:
\begin{itemize}
    \item $\text{Comm}^{\text{Agg}} = (\beta_0^{1,\text{Agg}}, \beta_1^{1,\text{Agg}}, \ldots, \beta_0^{\Psi,\text{Agg}}, \beta_1^{\Psi,\text{Agg}})$
    \item $\text{Res}_1^{\text{Agg}} = (g_1^{1,\text{Agg}}, h_1^{1,\text{Agg}}, \ldots, g_1^{\Psi,\text{Agg}}, h_1^{\Psi,\text{Agg}})$
    \item $\text{Res}_2^{\text{Agg}} = (f_{V_1}^{1,\text{Agg}}, \ldots, f_{V_\Psi}^{\Psi,\text{Agg}})$
\end{itemize}

For each round $j = 1, \ldots, \Psi$, the verifier checks whether the revealed witness $f_{V_j}^{j,\text{Agg}}$ satisfies the appropriate commitment verification equation. If $V_j^{\text{Agg}} = 0$, it verifies:
\[
\beta_0^{j,\text{Agg}} \stackrel{?}{=} \text{Commit}\left(
f_{V_j}^{j,\text{Agg}},\;
\delta_j f_{V_j}^{j,\text{Agg}} - g_1^{j,\text{Agg}},\;
\delta_j P(f_{V_j}^{j,\text{Agg}}) - h_1^{j,\text{Agg}}
\right)
\]

If $V_j^{\text{Agg}} = 1$, it verifies the following:
\begin{align}
\beta_1^{j,\text{Agg}} \stackrel{?}{=} \text{Commit}\big(& f_{V_j}^{j,\text{Agg}},\; \delta_j \left( s_{ID}^{\text{Agg}} - P(f_{V_j}^{j,\text{Agg}}) + P(0) \right) \nonumber \\
& {} - G(g_1^{j,\text{Agg}}, f_{V_j}^{j,\text{Agg}}) - h_1^{j,\text{Agg}} \big)
\end{align}

\begin{figure*}[!t]
\centering
\includegraphics[width=1\linewidth]{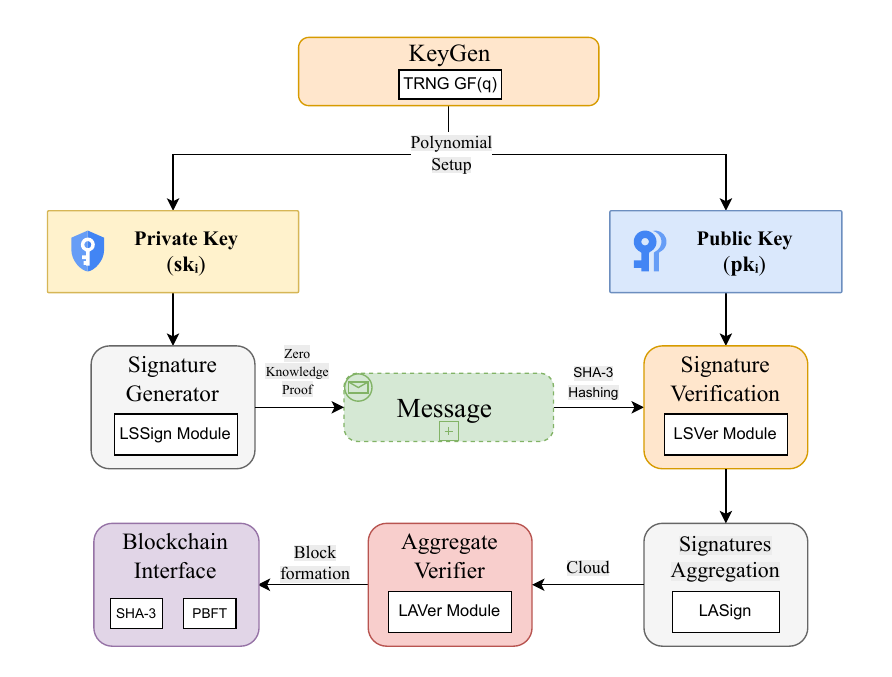}
\caption{PQ-MISS operational framework illustrating key generation, signing, verification, aggregation, and blockchain validation.}
\label{fig3}
\end{figure*}

If all $\Psi$ rounds pass their respective verification checks, the aggregate signature is accepted as valid. Otherwise, the verifier rejects the result. The modular structure in Figure~\ref {fig3} ensures separation between cryptographic operations and blockchain processes, enabling scalable deployment across UAV, fog, and cloud layers.

\newpage

\section{Applying PQ-MISS in IoT Applications}
\label{sec:application}
This section describes how the proposed PQ-MISS scheme is applied to real-world IoT-based scenarios, particularly in FANET (Flying Ad Hoc Network) environments. The network model comprises multiple entities: FANET devices, Fog Devices (FD), a Key Generation Center (KGC), and Cloud Servers (CS).

A set of $n_{cs}$ Cloud servers, denoted $\text{CS}_l$ where $l = 1, 2, \ldots, n_{cs}$, form a peer-to-peer Cloud server network, referred to as BC (Blockchain). This network is primarily responsible for validating and forming transaction blocks received from FDs and performing block mining using a consensus algorithm.

For each group of FANET devices $\text{SDp}_j$, the associated $\text{FD}_j$ is responsible for collecting data from the devices within its group. The KGC, also referred to as the Certificate Register (CR), is entrusted with the offline registration of both FANET devices and FDs associated with each group $\text{SDp}_j$.

The operational workflow begins with the secure registration of FANET devices and their corresponding fog nodes with the KGC before network deployment. Once deployed, each $\text{FD}_j$ collects timestamped, encrypted messages from its local FANET group $\text{SDp}_j$. These messages are then signed and aggregated at the fog level using PQ-MISS, where valid signatures are verified and compressed into a compact form. The verified aggregates are subsequently forwarded to $\text{CS}_l$ nodes, which package and mine the transactions into the blockchain using a consensus protocol. Finally, the system accommodates dynamic device deployment by allowing the real-time addition of new FANET devices and their re-registration through the KGC, ensuring that secure key updates and identity bindings are maintained.

All network entities are assumed to be time-synchronized, as recommended in prior studies~\cite{das2021igcacs, bera2022private, yu2022slap}. This temporal alignment ensures resistance against replay attacks by enabling accurate timestamp verification at every stage of communication and data processing.

\subsection{Registration}

The registration process in the proposed system applies to both each FANET device $\text{DR}_i$ and its associated fog node $\text{FD}_j$ within a group $\text{SDp}_j$. When $\text{SDp}_j$ contains $n_{\text{DR}}$ devices, the Control Room (CR), acting as the Key Generation Center (KGC), is responsible for completing their registration prior to deployment. 

In the case of \textbf{FANET device enrollment}, for each device $\text{DR}_i$, where $i = 1, 2, \ldots, n_{\text{DR}}$, the CR executes the KeyGen algorithm to generate a secret key $sk_i = u_{ID_i}$ and a corresponding public key $pk_i = S \circ F \circ T : \mathbb{F}_q^n \rightarrow \mathbb{F}_q^m$. A unique identifier $ID_{\text{DR}_i}$ is assigned to each device and, together with the key pair $(sk_i, pk_i)$, is preloaded into the device's secure memory. The CR then makes $pk_i$ publicly available.

For \textbf{fog node registration}, $\text{FD}_j$ first selects an identity $ID_j^{\text{Agg}}$ and securely transmits it to the CR. Acting as the KGC, the CR again runs the KeyGen algorithm to produce the fog node's secret key $sk_j^{\text{Agg}} = U_{ID_j}^{\text{Agg}}$ and its public key $pk_j^{\text{Agg}} = S \circ F \circ T$. These credentials are securely returned to $\text{FD}_j$ and stored in its secure database in the form $\{ID_j^{\text{Agg}}, (sk_j^{\text{Agg}}, pk_j^{\text{Agg}})\}$, while the public key $pk_j^{\text{Agg}}$ is also published for system-wide use.

\subsection{Data Collection}

Each FANET device $\text{DR}_i$ in a group $\text{SDp}_j$ performs data collection over a defined time window as follows:

\textbf{Step 1:} The device $\text{DR}_i$ generates a current timestamp $\text{CTS}_{\text{DR}_i}$ and prepares a message:
\[
m_i = \{ID_{\text{DR}_i}, ID_j^{\text{Agg}}, (\text{RTS}_{\text{Start}}, \text{RTS}_{\text{End}}), \text{Data}_{\text{DR}_i}, \text{CTS}_{\text{DR}_i}\}
\]
It then computes a signature $\text{SSig}_{m_i} = \text{LSSign}(sk_i, m_i)$ where $sk_i = U_{ID_i}$. Here, RTS refers to the start and end time range of the sensed data.

\textbf{Step 2:} $\text{DR}_i$ forms a transaction:
\[
\text{TX}_{\text{DR}_i} = \{m_i, \text{SSig}_{m_i}, pk_i\}
\]
and sends the message $\text{Msg}_{\text{DR}_i} = \{\text{TX}_{\text{DR}_i}\}$ to the corresponding $\text{FD}_j$ over a public channel.

\subsection{Data Aggregation}

The fog node $\text{FD}_j$ collects messages from its local FANET group and processes them as follows:

\textbf{Step 1:} Upon receiving $\text{Msg}_{\text{DR}_i}$ from device $i$ at time $\text{TS}_{\text{DR}_i}$, $\text{FD}_j$ verifies freshness using:
\[
|\text{CTS}_{\text{DR}_i} - \text{TS}_{\text{DR}_i}| < \Delta T
\]
If valid, the message is accepted. Otherwise, it is discarded.

\textbf{Step 2:} $\text{FD}_j$ runs the $\text{LSVer}$ algorithm on $m_i$ using public key $pk_i$. If the signature is valid, the transaction $\text{TX}_{\text{DR}_i}$ is accepted.

\textbf{Step 3:} For all valid transactions $\{\text{TX}_{\text{DR}_i}^1, \text{TX}_{\text{DR}_i}^2, \ldots, \text{TX}_{\text{DR}_i}^{t_n}\}$, the fog node executes $\text{LASign}$ using $sk_j^{\text{Agg}}$ to compute the aggregate signature:
\[
\text{ASig}_j^{\text{Agg}} = \text{LASign}(\{(\text{Sig}_{m_i}^k, m_i^k)\}_{k=1}^{t_n})
\]
It then sends the aggregation message:
\[
x_w = \{\text{TX}_{\text{DR}_i}^1, \ldots, \text{TX}_{\text{DR}_i}^{t_n}, \text{ASig}_j^{\text{Agg}}, pk_j^{\text{Agg}}\}
\]
to the Cloud server $\text{CS}_k$ over an open channel.

\subsection{Including Key Management during Addition and Verification of Blocks into Blockchain}

The proposed key management method aids in securing data collection during block creation in the private blockchain. In fog-enabled FANET applications, the data sensed from FANET devices are kept confidential and private to users. Sensitive data must not be publicly available. Furthermore, the blockchain ensures the security of data such that immutability, transparency, and decentralization of data are well-preserved.

\textbf{Notation Clarification:} In this section, we unify the terminology used across the fog and blockchain layers. The terms \textit{fog server (FSk)}, \textit{fog node}, and \textit{fog device (FDj)} are used interchangeably and refer to the same entity responsible for aggregating and forwarding data from FANET devices. For consistency, FSk corresponds to the fog device FDj introduced earlier. Similarly, GWNj and Grj denote the same FANET group previously defined as SDpj. The public and private keys of fog entities are represented uniformly as $(pk_{Agg_j}, sk_{Agg_j})$. All transaction notations $TX_w$ correspond to validated device transactions $TX_{DR_i}$ generated in the data aggregation phase. The proposed key management method aids in securing data collection during block creation.

The creation and operation of the blockchain involve the following steps:

\textbf{Step 1:} Initially, data securely transmitted from FANET devices ($\text{SD}_i$) to corresponding fog nodes ($\text{GWN}_j$) in group ($\text{Gr}_j$) are organized into several transactions ($\text{TX}_w$), where:
\[
\text{TX}_w = \{\text{Sensed Data},\; \text{ID of SD}_i \; (\text{RID}_{SD_i}),\; \text{Timestamp} \; (\text{TS}_{TX_w}) \}
\]
These transactions are securely forwarded to fog server $\text{FS}_k$ of $\text{GWN}_j$, encrypted based on public key $\text{Pub}_{FS_k}$.

\textbf{Step 2:} $\text{FS}_k$ gathers and encrypts transactions:
\[
\text{ETX}_w = E_{\text{Pub}_{FS_k}}[\text{TX}_w]
\]
and builds a Partial Block (ParBlk) from $n_t$ encrypted transactions ($\text{ETX}_w$). Then, Merkle Tree Root ($\text{MTR}_{\text{Blk}}$) is determined based on $n_t$ $\text{ETX}_w$ and Timestamp for Block formation ($\text{TS}_{\text{Blk}}$), type of CPS application (AppType), Block Owner ($\text{ID}_{FS_k}$), Owner's public key ($\text{Pub}_{FS_k}$), and ECDSA signature ($\text{Sig}_{\text{Blk}}$) on message:
\[
m = h(\text{TS}_{\text{Blk}} \parallel \text{MTR}_{\text{Blk}} \parallel \text{AppType} \parallel \text{ID}_{FS_k} \parallel \text{Pub}_{FS_k} \parallel \text{Sig}_{\text{Blk}})
\]
Encrypted transactions $(\text{ETX}_w, 1 \leq w \leq n_t)$ are signed using the private key of $\text{FS}_k$ ($\text{Pri}_{FS_k}$).

\begin{figure}
    \centering
    \includegraphics[width=1\linewidth]{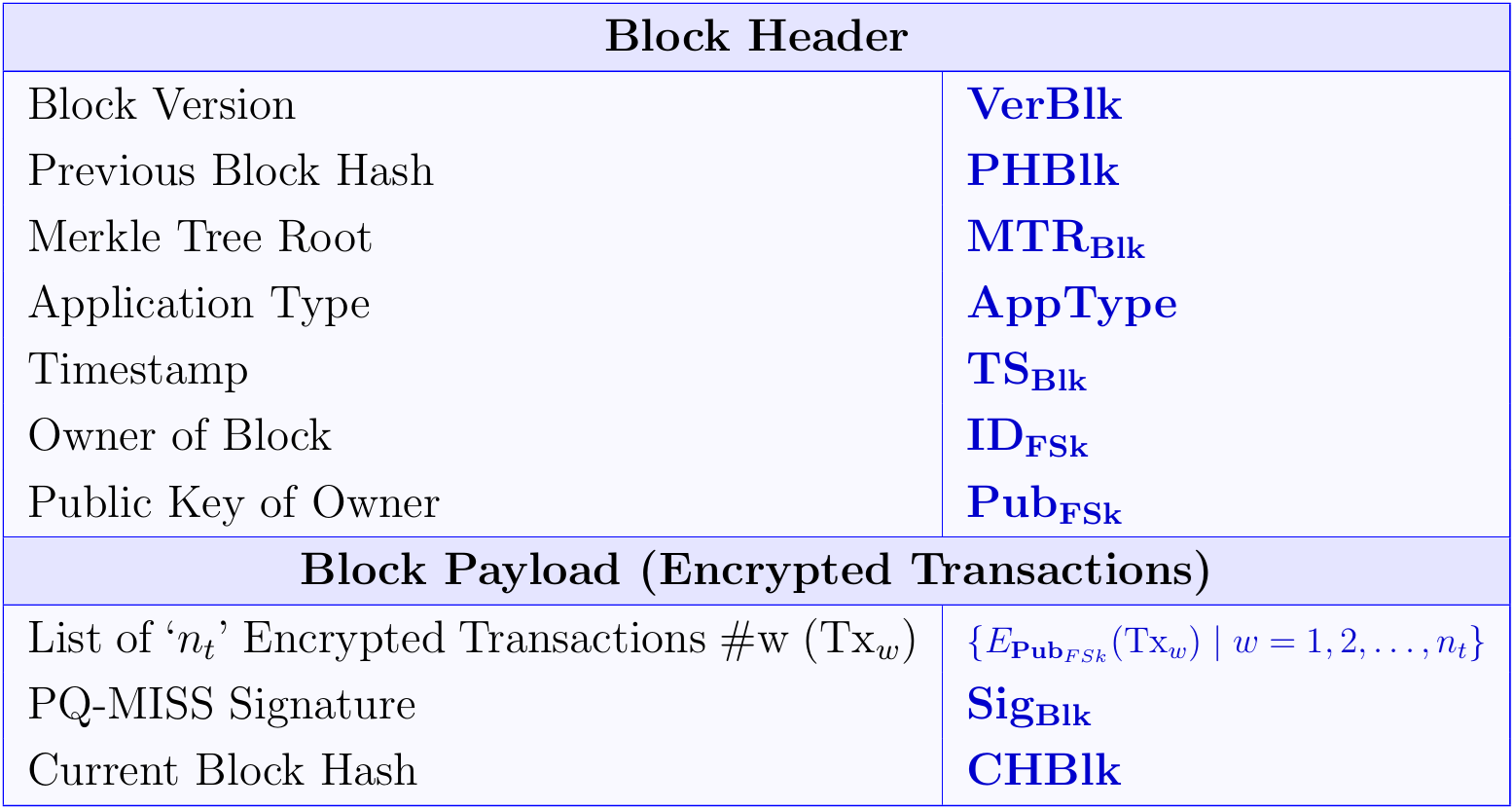}
    \caption{FullBlk For Various Transactions}
    \label{fig4}
\end{figure}

\textbf{Step 3:} Each $\text{FS}_k$ forwards the ParBlk to the associated CSs ($\text{CS}_l$) encrypted using the public key of $\text{CS}_l$ ($\text{Pub}_{CS_l}$). $\text{CS}_l$ includes Block Version (VerBlk), Previous Hash Block (PHBlk), and Current Hash Block (CHBlk) with the data from PHBlk and ParBlk. FullBlk is presented in Figure~\ref{fig4}.

\textbf{Step 4:} FullBlk is verified and included in BC's private blockchain using the Practical Byzantine Fault Tolerance (PBFT) consensus algorithm \cite{castro2002practical} in the P2P CS network. Algorithm~\ref{alg:verify-add-blocks} shows the verification and addition of blocks.

As shown in the network model, there are several companies that work together. Assume that fog servers along with FANET devices are in company A. CSs are in company B. AI-based Big Data analytics is performed by company C. The PBFT consensus scheme, which is voting-based, is used for verifying and including a FullBlk in the P2P CSs network. Incentives for fog servers and CSs for acting as miners and constructing blocks in the blockchain are not considered. Furthermore, ParBlk includes encrypted transactions $\text{ETX}_w = E_{\text{Pub}_{FS_k}}[\text{TX}_w]$, $\text{MTR}_{\text{Blk}}$, and ECDSA signature ($\text{Sig}_{\text{Blk}}$). ParBlk is sent to the corresponding CS ($\text{CS}_l$) which verifies the signature.

ParBlk is tested by checking whether MTR on encrypted transactions ($\text{ETX}_w$) matches the $\text{MTR}_{\text{Blk}}$ stored in the block. If valid, $\text{CS}_l$ confirms the integrity of ParBlk. Furthermore, $\text{ETX}_w$ and $\text{Sig}_{\text{Blk}}$ cannot be determined by any CSs, including $\text{CS}_l$ and an opponent, as it requires the private key ($\text{Pri}_{FS_k}$) of Fog Server ($\text{FS}_k$). $\text{CS}_l$ includes previous and current block hashes to ParBlk to convert it into FullBlk. The proposed mechanism focuses on issues related to the security and privacy of outsourced data.

\begin{algorithm}[H]
\caption{Verification and Addition of Blocks}
\label{alg:verify-add-blocks}
\begin{algorithmic}[1]
\Require $\textit{FullBlk}$;\; $\{(\mathrm{Pri}_{CS_\ell},\mathrm{Pub}_{CS_\ell})\}$;\;
         $n_{f}$  \Comment number of faulty nodes in P2P CS network
\Ensure Commitment and inclusion of $\textit{FullBlk}$ in every ledger

\State $L \gets$ leader chosen via Round Robin (RR) policy
\State $L$ produces timestamp $\mathrm{TS}_L$ and broadcasts
       $\langle \textit{FullBlk},\, E_{\mathrm{Pub}_{CS}}\!\left[ V_{\mathrm{req}}, \mathrm{TS}_L \right],\, \mathrm{TS}_L \rangle$
       to all $CS_\ell$ over a public channel
\ForAll{$CS_\ell$ \textbf{in parallel}}
    \State verify $\mathrm{TS}_L$
    \If{valid}
        \State $(V_{\mathrm{req}}, \mathrm{TS}_L') \gets
               D_{\mathrm{Pri}_{CS_\ell}}
               \!\left[ E_{\mathrm{Pub}_{CS}}\!\left[ V_{\mathrm{req}}, \mathrm{TS}_L \right] \right]$
        \State verify $\mathrm{TS}_L'$ against local time
        \State verify $\mathrm{MTR}_{Blk}$, $\mathrm{Sig}_{Blk}$, and $\mathrm{CHBlk}$ in $\textit{FullBlk}$
        \State send
               $\big\langle E_{\mathrm{Pub}_L}[\,V_{\mathrm{res}}, \mathrm{TS}_{CS_\ell}\,],\,
               \mathrm{TS}_{CS_\ell} \big\rangle$ to $L$
               \Comment $V_{\mathrm{res}}$: voting response
    \EndIf
\EndFor

\State $V_c \gets 0$ \Comment valid-vote counter
\ForAll{responses $\big\langle E_{\mathrm{Pub}_L}[\,V_{\mathrm{res}}, \mathrm{TS}_{CS_\ell}\,],\,
                    \mathrm{TS}_{CS_\ell} \big\rangle$}
    \State verify $\mathrm{TS}_{CS_\ell}$
    \State $V_{\mathrm{res}} \gets
           D_{\mathrm{Pri}_L}\!\left[ E_{\mathrm{Pub}_L}[\,V_{\mathrm{res}}, \mathrm{TS}_{CS_\ell}\,] \right]$
    \If{$V_{\mathrm{res}}$ is valid} \State $V_c \gets V_c + 1$ \EndIf
\EndFor

\If{$V_c > 2\,n_f + 1$}
    \State $L$ broadcasts \textbf{commit} for $\textit{FullBlk}$
    \State each $CS_\ell$ (including $L$) appends $\textit{FullBlk}$ to its ledger
\EndIf
\end{algorithmic}
\end{algorithm}

\subsection{Protected Prediction using Blockchain based on AI}
\label{sec:ai-blockchain}

Data from transactions stored in the blockchain are heterogeneous in nature, containing information about FANET device parameters such as location, speed, and altitude, as well as communication metrics including latency and signal strength. To enable effective AI/ML-based analysis, this data is preprocessed to construct structured datasets suitable for training and prediction tasks. The blockchain serves as a secure storage layer, ensuring that the data remains immutable, transparent, and accessible only to authorized components of the system. 

During simulation, FANET data is continuously collected and transmitted to blockchain nodes for secure storage. These data streams are subsequently accessed by AI models, which process them to generate predictions that inform decision-making within the network. This hybrid approach not only guarantees secure data handling but also leverages AI to provide meaningful insights while maintaining efficiency under the resource constraints of \texttt{ns-3}.

In the context of AI/ML analytics, transactional data recorded on the blockchain plays a critical role in determining system performance. Two distinct datasets are used: one for training the parameters of nodes, and another for testing the routing tables. In this work, a Long Short-Term Memory (LSTM) model is employed to predict the collected data, enhancing both the security and privacy of the hybrid FANET–fog–blockchain network. This integration of blockchain and AI thus supports resilient predictions and secure, intelligent decision-making.

\subsection{Dynamic Addition of Nodes}
\label{sec:dynamic-nodes}

To support scalability in FANET environments, the proposed system enables the dynamic addition of new devices after initial deployment. Consider the case of deploying a new FANET device $SD_i^{\text{new}}$ into an existing group $Gr_j$ managed by fog node $FN_j$. The Key Generation Center (KGC) first selects a unique identity $ID_{SD_i}^{\text{new}}$ and a corresponding registration timestamp $RTS_{SD_i}^{\text{new}}$. It then executes the KeyGen procedure to generate the device's secret–public key pair, where the private key is defined as $Pri_{SD_i}^{\text{new}} = U_{ID}^{\text{new}}$ and the public key is given by $Pub_{SD_i}^{\text{new}} = S \circ F \circ T : \mathbb{F}_q^n \rightarrow \mathbb{F}_q^m$.

The Control Room (CR), acting as the KGC, securely loads the tuple $\{Pri_{SD_i}^{\text{new}}, Pub_{SD_i}^{\text{new}}, ID_{SD_i}^{\text{new}}\}$ into the memory of $SD_i^{\text{new}}$ prior to deployment within $Gr_j$. Simultaneously, the public key $Pub_{SD_i}^{\text{new}}$ is published to the system directory, making the new device immediately recognizable and verifiable by other entities in the network. This procedure ensures that newly added nodes are securely integrated into the existing FANET–fog–blockchain ecosystem without compromising system integrity.

\section{Performance Evaluation}
\label{sec:evaluation}

To validate the effectiveness of the proposed scheme, extensive simulations are conducted on the NS-3.25 network simulator under a fog-enabled FANET scenario. The evaluation quantifies cryptographic overhead, scalability under message aggregation, and the efficiency of blockchain integration, with comparisons against two state-of-the-art schemes: MV-MSS \cite{srivastava2022blockchain} and LBAS \cite{bagchi2023public}. The simulation parameters are summarized in Table~\ref{tab:sim_params}.

\begin{table}[ht]
\small
\centering
\caption{Simulation parameters}
\label{tab:sim_params}
\renewcommand{\arraystretch}{1.15}
\setlength{\tabcolsep}{6pt}
\begin{tabularx}{\columnwidth}{@{}l X@{}}
\toprule
\textbf{Parameter} & \textbf{Value} \\
\midrule
Simulator & NS-3.25 \\
Simulation area (width × depth) & 1000 × 1000 m \\
Number of FANET devices & 50 \\
Number of fog devices & 10 \\
Number of cloud servers & 5 to 25 \\
Packet size & 100–10,000 bytes \\
External libraries used & NFL~\cite{nfl}, OpenSSL~\cite{openssl}, Crypto++~\cite{cryptopp} \\
Hash algorithm & SHA-3 (standard) \\
Protocol & PQ-MISS \\
Transmission range & 100 m \\
Simulation time & 100 s \\
\bottomrule
\end{tabularx}
\end{table}

 We take into account principal performance indicators that reflect cryptographic as well as network-level efficiency. (1) Signing time reflects the latency of generating digital signature at each UAV, showing the computational load on resource-constrained nodes. (2) Verification time reflects the processing delay faced by fog nodes or cloud servers when verifying individual or combined signatures, directly influencing real-time authentication and throughput. (3) Aggregate signing and verification times reflect the scalability of the scheme under the processing of multiple messages simultaneously at the fog layer, an indispensable parameter for swarm-level data agglomeration. (4) Total computation cost is the overall processing overhead including signature generation, verification, and blockchain processing across all nodes; this reflects the end-to-end efficiency of PQ-MISS's implementation within the blockchain-based fog architecture. These indicators collectively analyze the trade-off among security intensification, latency, and scalability, and illustrate the improved performance offered by the proposed PQ-MISS scheme vis-à-vis current post-quantum schemes like MV-MSS \cite{srivastava2022blockchain} and LBAS \cite{bagchi2023public}. The measured metrics collectively demonstrate the role of the multivariate structure and aggregate signature optimization in reducing computational complexity, thereby enhancing real-time performance in fog-enabled FANETs.

Figure~\ref{fig:times} presents the results for signing and verification times in both single-message and aggregated-message scenarios. For single messages (Figures~\ref{fig5a} and \ref{fig5b}), PQ-MISS achieves lower overhead than MV-MSS \cite{srivastava2022blockchain} and LBAS \cite{bagchi2023public} across all packet sizes. At the maximum packet size of 10,000 bytes, PQ-MISS reduces signing time by 35\% compared to MV-MSS \cite{srivastava2022blockchain} and 47.2\% compared to LBAS \cite{bagchi2023public}, while verification delay is reduced by 29.4\% and 54.8\%, respectively. These improvements underscore PQ-MISS's suitability for latency-sensitive UAV applications where rapid authentication is critical.

For aggregated operations (Figures~\ref{fig5c} and \ref{fig5d}), PQ-MISS maintains its advantage as the batch size increases. When 100 messages are aggregated, PQ-MISS requires only 0.65 s for signing, whereas MV-MSS \cite{srivastava2022blockchain} and LBAS \cite{bagchi2023public} require 0.83 s and 0.99 s, corresponding to improvements of 21.7\% and 34.3\%. Similar gains are observed for aggregated verification, confirming that PQ-MISS scales efficiently with aggregation and supports fog devices processing large volumes of UAV traffic.

\begingroup
\captionsetup[subfloat]{font=scriptsize}
\begin{figure}[!t]
\centering
\subfloat[Packet size vs. single message signing time]{%
  \includegraphics[width=0.5\linewidth]{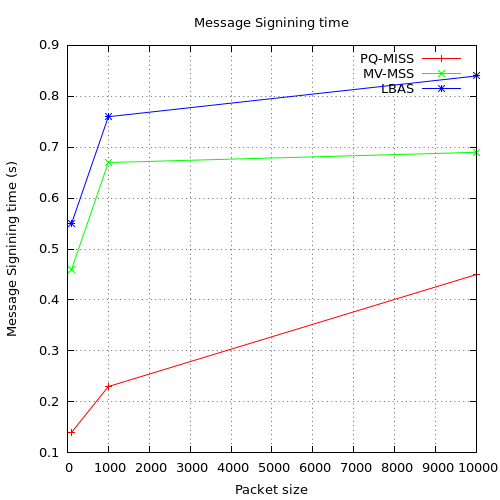}%
  \label{fig5a}}
\hfill
\subfloat[Packet size vs. single message verification time]{%
  \includegraphics[width=0.5\linewidth]{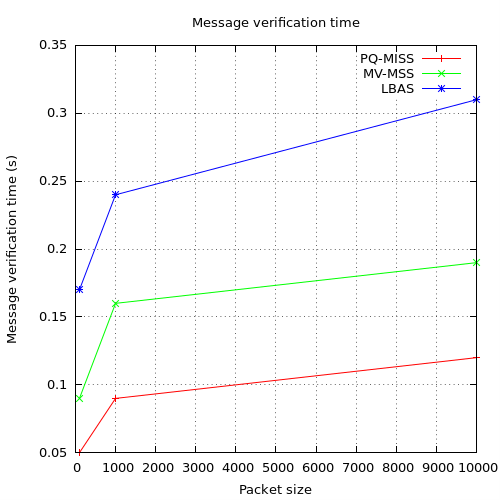}%
  \label{fig5b}}

\subfloat[Aggregated messages vs. aggregated signing time]{%
  \includegraphics[width=0.5\linewidth]{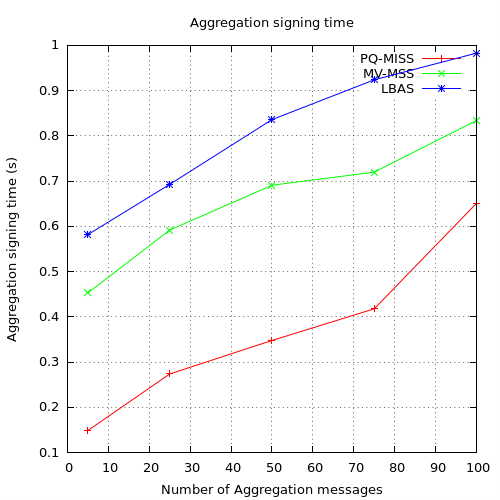}%
  \label{fig5c}}
\hfill
\subfloat[Aggregated messages vs. aggregated verification time]{%
  \includegraphics[width=0.5\linewidth]{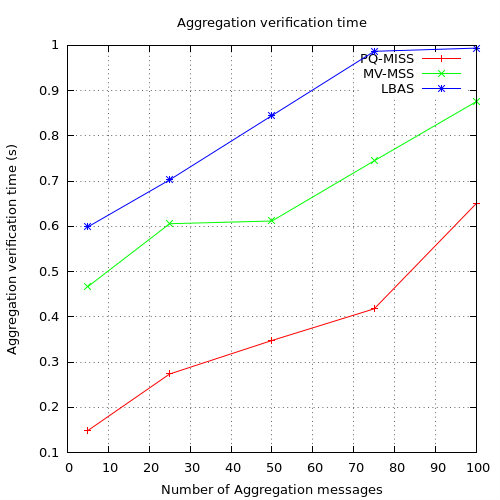}%
  \label{fig5d}}
\caption{Signing and verification times for single and aggregated messages.}
\label{fig:times}
\end{figure}
\endgroup

The computational efficiency of PQ-MISS in blockchain-enabled scenarios is shown in Figure~\ref{fig:total_cost}. Figure~\ref{fig6a} illustrates the total computation cost as the number of blockchain nodes increases from 5 to 25, with each block limited to 30 transactions and a maximum of 25 blocks overall. PQ-MISS achieves lower costs compared to the baseline schemes; at 25 nodes, the cost is 102 s compared to 109 s for MV-MSS \cite{srivastava2022blockchain} and 116 s for LBAS \cite{bagchi2023public}, reflecting improvements of 6.4\% and 12.1\%, respectively.

Figure~\ref{fig6b} evaluates total computation cost against the number of transactions, ranging from 30 to 80. PQ-MISS outperforms the baseline methods in this scenario as well. At 80 transactions, PQ-MISS incurs a cost of 64 s, while MV-MSS \cite{srivastava2022blockchain} and LBAS \cite{bagchi2023public} require 71 s and 82 s, respectively. This corresponds to reductions of 9.9\% relative to MV-MSS \cite{srivastava2022blockchain} and 21.9\% compared to LBAS \cite{bagchi2023public}. The results across all experiments confirm that PQ-MISS not only reduces computation costs but also scales more efficiently as network size and workload increase, making it a strong candidate for secure and resource-constrained UAV/FANET environments.

\begingroup
\captionsetup[subfloat]{font=scriptsize}
\begin{figure}[!t]
\centering
\subfloat[Number of blockchain nodes (P2P nodes) vs. total computation cost]{%
  \includegraphics[width=0.5\linewidth]{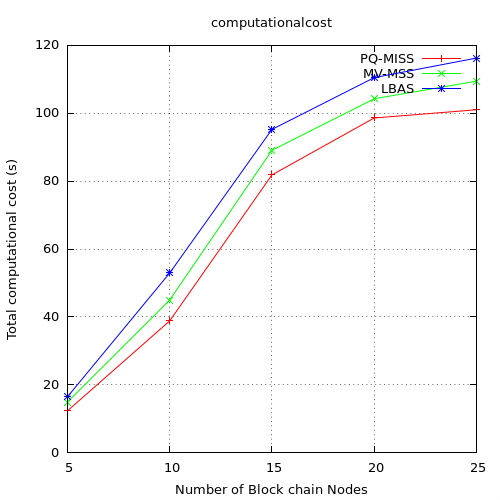}%
  \label{fig6a}}
\hfill
\subfloat[Number of transactions vs. total computation cost]{%
  \includegraphics[width=0.5\linewidth]{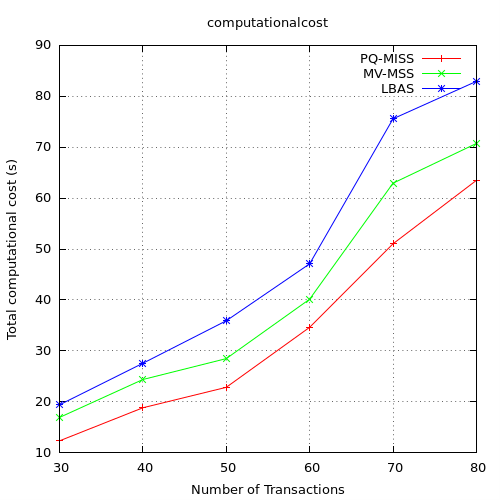}%
  \label{fig6b}}
\caption{Computational efficiency of PQ-MISS in blockchain-enabled scenarios.}
\label{fig:total_cost}
\end{figure}
\endgroup

The comprehensive evaluation demonstrates that PQ-MISS consistently outperforms existing schemes across multiple performance metrics, showing significant improvements in both individual signature operations and aggregate processing scenarios, while maintaining superior scalability in blockchain-integrated environments.

While the presented results demonstrate the efficiency and scalability of PQ-MISS, several limitations should be acknowledged. The simulations assumed a fixed mobility model and did not explicitly evaluate energy consumption, which is critical in UAV operations. Additionally, the experiments were conducted on a moderate-scale testbed (50 UAVs, 10 fog devices, and up to 25 blockchain nodes); larger deployments may introduce further communication and synchronization overhead. Furthermore, the integration of AI-assisted prediction components, discussed in Section~\ref{sec:application}, was not benchmarked here and will be explored in future work. Addressing these aspects represents an important direction for extending this study.

\section{Conclusion and future work}
\label{sec:conclusion}
In this work, we proposed a novel Post-Quantum Multivariate Digital Signature Scheme (PQ-MISS), which builds upon a secure MPKC signature scheme and a 5-Pass identification protocol. The scheme was tailored for fog-enabled FANET environments, where UAVs act as individual signers and fog devices (FDs) serve as aggregators. Aggregated messages and their corresponding signatures are verified by cloud servers operating within a peer-to-peer network, which subsequently form and validate blocks on the blockchain. This integration ensures immutability, decentralization, and transparency across the system.

Comprehensive security analysis demonstrates that PQ-MISS is resistant to a broad spectrum of threats, including quantum-capable adversaries. Performance evaluation using NS-3 simulations confirmed the efficiency of our design, measuring computational costs for both individual and aggregate signature generation and verification at the UAV, FD, and cloud server levels. The results validate the practical feasibility of the proposed approach in resource-constrained environments.

In future work, we plan to extend PQ-MISS to incorporate adaptive mobility and energy-aware models, perform large-scale experimental validations, and fully integrate and benchmark the AI-driven prediction modules for real-time anomaly detection and trust management. These enhancements will complete the transition toward a fully intelligent, quantum-resilient, and self-optimizing FANET framework.

\section*{CRediT authorship contribution statement}

\textbf{Sufian Al Majmaiea}: Conceptualization, Methodology, Software, Formal analysis, Investigation, Writing – Original Draft, Visualization. \textbf{Ghazal Ghajari}: Supervision, Validation, Writing – Review \& Editing, Resources. \textbf{Niraj Prasad Bhatta}: Software, Validation, Writing – review and editing, Data curation. \textbf{Fathi Amsaad}: Project administration, Funding acquisition, Review \& Editing, Technical guidance.

\section*{Acknowledgment}

The authors would like to express their heartfelt gratitude for the financial and technical support provided under the Assured and Trusted Digital Microelectronics Ecosystem (ADMETE) grant, BAA-FA8650-18-S-1201, funded by the Air Force Research Laboratory (AFRL). Partial support for this work was also received through funding to Wright State University, Dayton, Ohio, USA by the National Security Agency (NSA). This project adhered to CAGE Number: 4B991 and DUNS Number: 047814256.

\section*{Declaration of competing interest}

The authors declare that they have no known competing financial interests or personal relationships that could have appeared to influence the work reported in this paper.

\section*{Data availability} 

No data was used for the research described in the article 

\bibliography{references}

@article{mosenia2016comprehensive,
  title        = {A comprehensive study of security of {Internet of Things}},
  author       = {Mosenia, Arsalan and Jha, Niraj K},
  journal      = {IEEE Transactions on Emerging Topics in Computing},
  volume       = {5},
  number       = {4},
  pages        = {586--602},
  year         = {2016}
}

@article{boyes2018industrial,
  title        = {The {Industrial Internet of Things} ({IIoT}): An analysis framework},
  author       = {Boyes, Hugh and Hallaq, Bil and Cunningham, Joe and Watson, Tim},
  journal      = {Computers in Industry},
  volume       = {101},
  pages        = {1--12},
  year         = {2018}
}

@article{zhang2014sybil,
  title        = {Sybil attacks and their defenses in the {Internet of Things}},
  author       = {Zhang, Kuan and Liang, Xiaohui and Lu, Rongxing and Shen, Xuemin},
  journal      = {IEEE Internet of Things Journal},
  volume       = {1},
  number       = {5},
  pages        = {372--383},
  year         = {2014}
}

@article{bernstein2017post,
  title        = {Post-quantum cryptography},
  author       = {Bernstein, Daniel J and Lange, Tanja},
  journal      = {Nature},
  volume       = {549},
  number       = {7671},
  pages        = {188--194},
  year         = {2017}
}

@inproceedings{ding2005rainbow,
  title        = {Rainbow, a new multivariable polynomial signature scheme},
  author       = {Ding, Jintai and Schmidt, Dieter},
  booktitle    = {International Conference on Applied Cryptography and Network Security},
  pages        = {164--175},
  year         = {2005},
  organization = {Springer}
}

@inproceedings{sakumoto2011provable,
  title        = {On provable security of {UOV} and {HFE} signature schemes against chosen-message attack},
  author       = {Sakumoto, Koichi and Shirai, Taizo and Hiwatari, Harunaga},
  booktitle    = {Post-Quantum Cryptography: 4th International Workshop (PQCrypto 2011), Taipei, Taiwan, Nov 29--Dec 2, 2011. Proceedings},
  pages        = {68--82},
  year         = {2011},
  organization = {Springer}
}

@article{chen2019identity,
  title        = {Identity-based signature schemes for multivariate public key cryptosystems},
  author       = {Chen, Jiahui and Ling, Jie and Ning, Jianting and Ding, Jintai},
  journal      = {The Computer Journal},
  volume       = {62},
  number       = {8},
  pages        = {1132--1147},
  year         = {2019}
}

@inproceedings{shen2013ibuov,
  title        = {{IBUOV}, a provably secure identity-based {UOV} signature scheme},
  author       = {Shen, Wuqiang and Tang, Shaohua and Xu, Lingling},
  booktitle    = {2013 IEEE 16th International Conference on Computational Science and Engineering},
  pages        = {388--395},
  year         = {2013},
  organization = {IEEE}
}

@article{luyen2019improved,
  title        = {An improved identity-based multivariate signature scheme based on {Rainbow}},
  author       = {Luyen, Le Van},
  journal      = {Cryptography},
  volume       = {3},
  number       = {1},
  pages        = {8},
  year         = {2019}
}

@inproceedings{dorri2017blockchain,
  title        = {Blockchain for {IoT} security and privacy: The case study of a smart home},
  author       = {Dorri, Ali and Kanhere, Salil S and Jurdak, Raja and Gauravaram, Praveen},
  booktitle    = {2017 IEEE International Conference on Pervasive Computing and Communications Workshops (PerCom Workshops)},
  pages        = {618--623},
  year         = {2017},
  organization = {IEEE}
}

@article{bera2022private,
  title        = {Private blockchain-envisioned drones-assisted authentication scheme in {IoT}-enabled agricultural environment},
  author       = {Bera, Basudeb and Vangala, Anusha and Das, Ashok Kumar and Lorenz, Pascal and Khan, Muhammad Khurram},
  journal      = {Computer Standards \& Interfaces},
  volume       = {80},
  pages        = {103567},
  year         = {2022}
}

@article{yu2022slap,
  title        = {{SLAP}-{IoD}: Secure and lightweight authentication protocol using physical unclonable functions for Internet of drones in smart city environments},
  author       = {Yu, Sungjin and Das, Ashok Kumar and Park, Youngho and Lorenz, Pascal},
  journal      = {IEEE Transactions on Vehicular Technology},
  volume       = {71},
  number       = {10},
  pages        = {10374--10388},
  year         = {2022}
}

@article{salem2024advancing,
  title        = {Advancing cybersecurity: a comprehensive review of {AI}-driven detection techniques},
  author       = {Salem, Aya H and Azzam, Safaa M and Emam, O. E. and Abohany, Amr A},
  journal      = {Journal of Big Data},
  volume       = {11},
  number       = {1},
  pages        = {105},
  year         = {2024}
}

@article{okdem2024artificial,
  title        = {Artificial intelligence in cybersecurity: A review and a case study},
  author       = {Okdem, Selcuk and Okdem, Sema},
  journal      = {Applied Sciences},
  volume       = {14},
  number       = {22},
  pages        = {10487},
  year         = {2024}
}

@article{ullah2024aicyber,
  title        = {{AICyber-Chain}: Combining {AI} and blockchain for improved cybersecurity},
  author       = {Ullah, Zia and Waheed, Abdul and Mohmand, Muhammad Ismail and Basar, Sadia and Zareei, Mahdi and Granda, Fausto},
  journal      = {IEEE Access},
  year         = {2024}
}

@article{hulsing2016mq,
  title        = {From 5-pass {MQ}-based identification to {MQ}-based signatures},
  author       = {H{\"u}lsing, Andreas and Rijneveld, Joost and Samardjiska, Simona and Schwabe, Peter},
  journal      = {IACR Cryptology ePrint Archive},
  volume       = {2016},
  pages        = {708},
  year         = {2016}
}

@inproceedings{paterson2006efficient,
  title        = {Efficient identity-based signatures secure in the standard model},
  author       = {Paterson, Kenneth G and Schuldt, Jacob C. N.},
  booktitle    = {Australasian Conference on Information Security and Privacy},
  pages        = {207--222},
  year         = {2006},
  organization = {Springer}
}

@article{das2021igcacs,
  title        = {{iGCACS}-{IoD}: An improved certificate-enabled generic access control scheme for Internet of drones deployment},
  author       = {Das, Ashok Kumar and Bera, Basudeb and Wazid, Mohammad and Jamal, Sajjad Shaukat and Park, Youngho},
  journal      = {IEEE Access},
  volume       = {9},
  pages        = {87024--87048},
  year         = {2021}
}

@inproceedings{zhang2022recurrent,
  title        = {Recurrent {LSTM}-based {UAV} trajectory prediction with {ADS}-{B} information},
  author       = {Zhang, Yifan and Jia, Ziye and Dong, Chao and Liu, Yuntian and Zhang, Lei and Wu, Qihui},
  booktitle    = {GLOBECOM 2022 -- IEEE Global Communications Conference},
  pages        = {1--6},
  year         = {2022},
  organization = {IEEE}
}

@article{xu2001hm,
  title        = {{HM} revisits the Tower of Hanoi puzzle},
  author       = {Xu, Yaoda and Corkin, Suzanne},
  journal      = {Neuropsychology},
  volume       = {15},
  number       = {1},
  pages        = {69},
  year         = {2001}
}

@article{goldwasser1988digital,
  title        = {A digital signature scheme secure against adaptive chosen-message attacks},
  author       = {Goldwasser, Shafi and Micali, Silvio and Rivest, Ronald L},
  journal      = {SIAM Journal on Computing},
  volume       = {17},
  number       = {2},
  pages        = {281--308},
  year         = {1988}
}

@article{castro2002practical,
  title        = {Practical {Byzantine} fault tolerance and proactive recovery},
  author       = {Castro, Miguel and Liskov, Barbara},
  journal      = {ACM Transactions on Computer Systems},
  volume       = {20},
  number       = {4},
  pages        = {398--461},
  year         = {2002}
}

@misc{nfl,
  author       = {Quarkslab},
  title        = {{NFL}lib -- Number Theoretic Transform Library},
  year         = {2016},
  howpublished = {\url{https://github.com/quarkslab/NFLlib}},
  note         = {Accessed: 2025-06-24}
}

@misc{openssl,
  author       = {{The OpenSSL Project}},
  title        = {OpenSSL 1.1.0 Source Archive},
  year         = {2016},
  howpublished = {\url{https://ftp.openssl.org/source/old/1.1.0/openssl-1.1.0.tar.gz}},
  note         = {Accessed: 2025-06-24}
}

@misc{cryptopp,
  author       = {Wei Dai and Crypto++ Community},
  title        = {Crypto++ Library},
  year         = {1995--2025},
  howpublished = {\url{https://www.cryptopp.com/}},
  note         = {Accessed: 2025-06-24}
}

@inproceedings{holcomb2021pqfabric,
  title        = {{PQFabric}: A permissioned blockchain secure from both classical and quantum attacks},
  author       = {Holcomb, Amelia and Pereira, Geovandro and Das, Bhargav and Mosca, Michele},
  booktitle    = {2021 IEEE International Conference on Blockchain and Cryptocurrency (ICBC)},
  pages        = {1--9},
  year         = {2021},
  organization = {IEEE}
}

@article{kim2024post,
  title        = {Post-quantum delegated {Proof of Luck} for blockchain consensus algorithm},
  author       = {Kim, Hyunjun and Kim, Wonwoong and Kang, Yeajun and Kim, Hyunji and Seo, Hwajeong},
  journal      = {Applied Sciences},
  volume       = {14},
  number       = {18},
  pages        = {8394},
  year         = {2024}
}

@article{zhang2024post,
  title        = {Post-quantum secure identity-based signature scheme with lattice assumption for {Internet of Things} networks},
  author       = {Zhang, Yang and Tang, Yu and Li, Chaoyang and Zhang, Hua and Ahmad, Haseeb},
  journal      = {Sensors},
  volume       = {24},
  number       = {13},
  pages        = {4188},
  year         = {2024}
}

@article{prajapat2024designing,
  title        = {Designing high-performance identity-based quantum signature protocol with strong security},
  author       = {Prajapat, Sunil and Kumar, Pankaj and Kumar, Sandeep and Das, Ashok Kumar and Shetty, Sachin and Hossain, M. Shamim},
  journal      = {IEEE Access},
  volume       = {12},
  pages        = {14647--14658},
  year         = {2024}
}

@inproceedings{dong2023blockchain,
  title        = {Blockchain-based identity authentication oriented to multi-cluster {UAV} networking},
  author       = {Dong, Zesong and Tong, Wei and Zhang, Zhiwei and Li, Jian and Yang, Weidong and Shen, Yulong},
  booktitle    = {2023 IEEE International Conference on Blockchain (Blockchain)},
  pages        = {68--73},
  year         = {2023},
  organization = {IEEE}
}

@article{srivastava2022blockchain,
  title={Blockchain-envisioned provably secure multivariate identity-based multi-signature scheme for Internet of Vehicles environment},
  author={Srivastava, Vikas and Debnath, Sumit Kumar and Bera, Basudeb and Das, Ashok Kumar and Park, Youngho and Lorenz, Pascal},
  journal={IEEE Transactions on Vehicular Technology},
  volume={71},
  number={9},
  pages={9853--9867},
  year={2022},
  publisher={IEEE}
}

@article{bagchi2023public,
  title={Public blockchain-envisioned security scheme using post quantum lattice-based aggregate signature for internet of drones applications},
  author={Bagchi, Prithwi and Maheshwari, Raj and Bera, Basudeb and Das, Ashok Kumar and Park, Youngho and Lorenz, Pascal and Yau, David KY},
  journal={IEEE Transactions on Vehicular Technology},
  volume={72},
  number={8},
  pages={10393--10408},
  year={2023},
  publisher={IEEE}
}
\end{document}